\documentclass[12pt]{article}

\ifx\pdfoutput\undefined
\usepackage[dvips,bookmarks]{hyperref}
\else
\usepackage{hyperref}
\fi
\hypersetup{colorlinks=false,bookmarksopen,bookmarksnumbered,citecolor=blue,
   pdfstartview=FitH}

\usepackage{latexsym}

\usepackage{amssymb,amsfonts,amsmath}
\usepackage{graphicx} 
\usepackage{indentfirst}

 \usepackage{bbm}

\parskip=\medskipamount

\newcommand {\cC}{{\cal C}}
\newcommand {\cD}{{\cal D}}
\newcommand {\cE}{{\cal E}}

\newcommand {\cJ}{{\cal J}}
\newcommand {\cK}{{\cal K}}
\newcommand {\cL}{{\cal L}}
\newcommand {\cM}{{\cal M}}

\newcommand {\cN}{{\cal N}}

\newcommand {\cR}{{\cal R}}
\newcommand {\cS}{{\cal S}}

\newcommand {\cV}{{\cal V}}

\newcommand {\cY}{{\cal Y}}

\newcommand{\bG}{{\bf G}}

\newcommand{\bZ}{{\bf Z}}
\def\a{\alpha}

\def\b{\beta}
\def\c{\chi}
\def\d{\delta}

\def\f{\phi}
\def\g{\gamma}
\def\G{\Gamma}

\def\j{\psi}
\def\k{\kappa}
\def\l{\lambda}
\def\m{\mu}

\def\o{\omega}

\def\q{\theta}
\def\r{\rho}
\def\s{\sigma}
\def\t{\tau}

\def\x{\xi}
\def\z{\zeta}
\def\D{\Delta}
\def\F{\Phi}
\def\J{\Psi}
\def\L{\Lambda}
\def\O{\Omega}

\def\S{\Sigma}
\def\U{\Upsilon}

\def\rd{{\rm d}}
\def\ri{{\rm i}}
\def\re{{\rm e}}

\newcommand{\ad}{{\dot{\alpha}}}                           
\newcommand{\bd}{{\dot{\beta}}}                            
\newcommand{\ve}{\varepsilon}                            
\newcommand{\cDB}{{\bar\cD}}                            
\newcommand{\DB}{\bar{D}}

\newcommand{\ab}{{\a\b}}

\newcommand{\pa}{\partial}                           
\newcommand{\hf}{\frac12}

\newcommand{\vf}{\varphi}

\newcommand{\be}{\begin{equation}}
\newcommand{\ee}{\end{equation}}
\newcommand{\bea}{\begin{eqnarray}}
\newcommand{\eea}{\end{eqnarray}}
\newcommand{\non}{\nonumber}
\newcommand{\ba}{\begin{array}}
\newcommand{\ea}{\end{array}}

\newcommand{\1}{\underline{1}}

\newcommand{\mun}{\underline{m}}
\newcommand{\bm}[1]{\mbox{\boldmath$#1$}}

\def\double #1{#1{\hbox{\kern-2pt $#1$}}}

\newcommand{\dd}{{\dot\d}}

\newcommand{\teb}{{\bar{\theta}}}

\newcommand{\bsubeq}{\begin{subequations}}
\newcommand{\esubeq}{\end{subequations}}

\newcommand{\qb}{{\bar{\theta}}}

\newcommand{\Fb}{{\bar{\F}}}

\newcommand{\bfD}{{\bf D}}
\newcommand{\bfDB}{{\bar{\bfD}}}

\newcommand{\fb}{{\bar{\phi}}}

\newcommand{\de}{{\nabla}}
\newcommand{\deb}{{\bar{\nabla}}}

\newcommand{\mub}{{\bar{\mu}}}
\newcommand{\lb}{{\bar{\l}}}

\begin{document}
\begin{titlepage}
\begin{flushright}
UUITP-25/11\\
September 2011\\
\end{flushright}

\begin{center}
{\Large \bf 
Three-dimensional  $\bm{\cN = 2}$  (AdS) supergravity \\and associated supercurrents }
\end{center}

\begin{center}

{\bf
Sergei M. Kuzenko\footnote{kuzenko@cyllene.uwa.edu.au}${}^{a}$
and
Gabriele Tartaglino-Mazzucchelli\footnote{gabriele.tartaglino-mazzucchelli@physics.uu.se}${}^{b}$
} \\
\vspace{5mm}

\footnotesize{
${}^{a}${\it School of Physics M013, The University of Western Australia\\
35 Stirling Highway, Crawley W.A. 6009, Australia}}  
~\\
\vspace{2mm}

\footnotesize{
${}^{b}${\it Theoretical Physics, Department of Physics and Astronomy,
Uppsala University \\ 
Box 516, SE-751 20 Uppsala, Sweden}
}
\vspace{2mm}

\end{center}

\begin{abstract}
\baselineskip=14pt
Long ago,  Ach\'ucarro and Townsend discovered that in three dimensions (3D)
$\cN$-extended anti-de Sitter (AdS) supergravity exists in several incarnations, 
which were called
the  $(p,q)$ AdS supergravity theories  with non-negative integers $p \geq q$ such that 
$\cN=p+q$.
Using the superspace approach to 3D $\cN$-extended supergravity 
developed in arXiv:1101.4013, we present three superfield formulations 
for $\cN=2$ supergravity that allow for well defined  cosmological terms and supersymmetric 
AdS solutions. The conformal compensators corresponding to these theories are respectively:
(i) a chiral scalar multiplet; (ii) a vector multiplet; and (iii) an improved complex linear multiplet.
The theories corresponding to (i) and 
(iii)   are shown to provide two dually equivalent realizations of the (1,1) AdS supergravity,
while (ii)  describes the (2,0) AdS supergravity. 
We associate with each supergravity formulation, with and without  a cosmological term,
a consistent supercurrent multiplet. The supercurrents in the (1,1) and (2,0) AdS backgrounds
are derived for the first time. We elaborate on rigid supersymmetric theories in (1,1) and 
(2,0) AdS superspaces.
\end{abstract}

\vfill
\end{titlepage}

\newpage
\renewcommand{\thefootnote}{\arabic{footnote}}
\setcounter{footnote}{0}

\tableofcontents{}
\vspace{1cm}
\bigskip\hrule


\section{Introduction}
\setcounter{equation}{0}

A great many $\cN=2$ supersymmetric theories in three dimensions (3D) can be obtained 
by dimensional reduction from 4D $\cN=1$ supersymmetric systems. 
Of particular interest, however,  are those theories which do not allow for such a construction.
They are characterized by purely 3D phenomena\footnote{The physical
phenomena specific to three dimensions include the existence of  
real mass terms generated by a  central charge. However, such mass terms can be obtained 
by dimensionally reducing a 4D $\cN=1$ system of chiral multiplets coupled to a certain background 
vector multiplet \cite{Siegel80}. 
In the context of supersymmetric nonlinear $\s$-models,
such mass terms were introduced for the first time in  
two dimensions \cite{A-GF} by using the Scherk-Schwarz mechanism for dimensional reduction \cite{SS}.}
such as Chern-Simons couplings that are ubiquitous in three dimensions, both in pure
gravity \cite{DJT1,DJT2,Witten,HW} and 
supergravity \cite{vN,RvN,AT,LR89,AT2,HIPT}.
A non-trivial  example 
of $\cN=2$ supersymmetric theories with Chern-Simons terms
is the so-called $(2,0)$ 
anti-de Sitter (AdS) supergravity studied in \cite{AT,IT}.
More specifically, Ach\'ucarro and Townsend \cite{AT} discovered that in three dimensions 
$\cN$-extended AdS supergravity exists in several incarnations. 
These were called the  $(p,q)$ AdS supergravity theories  
where the  non-negative integers $p \geq q$ are such that 
$\cN=p+q$.   It was shown in \cite{AT} that these theories  
are naturally  associated with the 3D AdS supergroups
$\rm OSp (p|2; {\mathbb R} ) \times  OSp (q|2; {\mathbb R} )$.
The $(0,0)$ theory is simply 3D gravity with  a negative cosmological term. 
The $(1,0)$ theory  coincides with the $\cN=1$ AdS supergravity first presented in \cite{GGRS}. 
In the simplest extended case $\cN=2$,
 two different AdS supergravity theories emerge,  $(1,1)$ and $(2,0)$,
of which the former may be obtained by dimensional reduction from 4D $\cN=1$ AdS supergravity, 
while the latter is truly novel. It turns out that  $(1,1)$ and $(2,0)$ AdS supergravity theories possess
drastically different matter couplings. At the component level, certain matter couplings in 
$(2,0)$ AdS supergravity  were studied in \cite{IT}. To the best of our knowledge, 
a superspace analysis of such problems 
has not yet appeared in the literature (a special off-shell version of 3D $\cN=2$ Poincar\'e supergravity 
was presented in \cite{NG}).
One of the goals of this paper is to fill this gap.

A robust approach to engineering Poincar\'e supergravity theories in diverse dimensions is to describe them  as 
conformal supergravity coupled to certain compensating supermultiplet(s) \cite{KT}.
The same approach is clearly suitable to construct AdS supergravity models.
In the case of 3D $\cN$-extended conformal supergravity, conventional constraints 
on the superspace torsion were proposed in \cite{HIPT}, and some of their implications were also analyzed. 
Starting from these constraints,
in our recent work \cite{KLT-M}
the superspace geometry of 3D ${\cal N}$-extended conformal supergravity was
developed\footnote{The cases of $\cN=8$ and $\cN=16$ conformal supergravity theories 
 have been worked out in 
\cite{Howe:2004ib,CGN} and \cite{GH} respectively.} 
and then applied to construct general off-shell supergravity-matter couplings
for the cases ${\cal N} \leq 4$. 
In the present paper we make use of the approach of \cite{KLT-M}
in order to elaborate upon the case $\cN=2$.
The main goals of this work are to study (i) the $(1,1)$ and $(2,0)$ AdS supergravity theories; 
and (ii) supersymmetric field theory in $(1,1)$ and $(2,0)$ AdS superspaces, including 
a thorough analysis of the consistent supercurrent multiplets corresponding to the two types
of 3D $\cN=2$ AdS supersymmetry.

${}$From the point of view of Poincar\'e supergravity, the 3D $\cN=2$ and 4D $\cN=1$ theories are 
very similar. A non-trivial difference between them proves to emerge only in the AdS case. 
So let us first recall some general facts about the known off-shell versions of 4D $\cN=1$ supergravity 
(see \cite{GGRS,Ideas} for reviews), each of which can be realized as conformal supergravity coupled
to a compensator \cite{Howe,GGRS}. 
There exist three off-shell formulations of 4D $\cN=1$ Poincar\'e supergravity which are: (i) the old minimal 
($n=-1/3$) \cite{old} reviewed in \cite{WB}; the new minimal ($n=0$) \cite{new}; 
and (iii) the non-minimal ($n\neq -1/3, 0$) \cite{non-min,SG}.\footnote{The off-shell supergravity versions
are traditionally labelled by the real parameter $n$  introduced by Gates  and Siegel \cite{SG}.}
In the conformal supergravity setting, they differ by the choice of compensator, which is respectively:
(i) a chiral scalar multiplet;
(ii) a massless tensor multiplet; or (iii) a non-minimal scalar multiplet described by 
a complex linear scalar and its conjugate. For a long time it was believed \cite{GGRS} that only the old minimal 
formulation is suitable to realize AdS supergravity by adding an appropriate cosmological term to the 
supergravity action (see \cite{GGRS,Ideas} for reviews). Recently it has been shown \cite{BKdual}
that a certain version  of non-minimal supergravity, $n=-1$, is equally suitable to describe 
AdS supergravity. However, this is achieved not by 
 adding a cosmological term to the supergravity action, as in the $n=-1/3$ case, 
 but instead by deforming the complex linear constraint obeyed by the compensator.
The minimal and the non-minimal formulations of AdS supergravity are then dually equivalent \cite{BKdual}. 
As to the new minimal formulation of $\cN=1$ supergravity in four dimensions, $n=0$, it {\it cannot} be used to describe  AdS supergravity.

As regards 3D  $\cN=2$ Poincar\'e supergravity, it also allows three different off-shell formulations
\cite{KLT-M} which are associated with the  following choices of conformal compensator:
(i) a chiral scalar multiplet; (ii) a massless vector multiplet; and (iii) a non-minimal scalar multiplet 
described by  a complex linear scalar $\S$  and its conjugate. They are 3D analogues  of the old minimal, 
new minimal and non-minimal supergravity theories in four dimensions, respectively. 
The 3D supergravity versions (i) and (ii) will be called Type I minimal and Type II minimal, respectively, 
in what follows. As  shown in \cite{KLT-M}, 
the 3D non-minimal theory is naturally parametrized by the super-Weyl weight of $\S$, denoted  $w$, 
which proves to be  related to the 4D  Siegel-Gates parameter $n$ as follows:
\bea
n= \frac{1-w}{3w+1}~.
\label{1.1}
\eea
As in four dimensions, the Type I minimal and the $w=-1$ (or $n=-1$) non-minimal formulations 
can be used to describe AdS supergravity by modifying the supergravity action (in the Type I case)
or deforming the complex linear constraint (in the non-minimal case). 
The two realizations turn out to be dually equivalent 
and lead to the same $(1,1)$ AdS supergravity. Unlike the situation in 
four dimensions, the Type II theory can also be used to describe AdS supergravity, for now 
a cosmological term can be realized as the supersymmetric Chern-Simons term associated 
with the compensating vector multiplet. Adding such a cosmological term to the Type II supergravity action 
provides a superspace description of $(2,0)$ AdS supergravity! 

For both (1,1) and (2,0) AdS supergravity theories, the equations of motion prove to require the superspace 
geometry to have constant torsion and curvature. In the case of 
Type I AdS supergravity, the on-shell geometry is described by covariant derivatives 
\bea
\de_A = (\de_a , \de_\a, \bar \de^\a)= {E}_A{}^M \pa_M +\hf \O_A{}^{cd}\cM_{cd}
\label{1.2}
\eea
obeying the following algebra
\bsubeq \label{11AdSsuperspace}
\bea
\{\de_\a,\de_\b\}
&=&
-4\bar{\mu}\cM_{\a\b}
~,~~~
 \{\deb_\a,\deb_\b\}
=
4\mu\cM_{\a\b}
~,~~~
\{\de_\a,\deb_\b\}
=
-2\ri\de_{\a\b}~, 
\label{AdS_(1,1)_algebra_1}
\\
{[}\de_{\a\b},\de_\g{]}
&=&
-2\ri\bar{\mu}\ve_{\g(\a}\deb_{\b)}
~,\qquad
{[}\de_{\a\b},\deb_\g{]}
=
2\ri\mu\ve_{\g(\a}\de_{\b)}
~,
\label{AdS_(1,1)_algebra_3/2}
 \\
{[}\de_a,\de_b]{}
&=& -4 \bar{\mu}\mu \cM_{ab} ~,
\label{AdS_(1,1)_algebra_2}
\eea
\esubeq
with $\m$ a constant complex parameter,  and $\cM_{ab}=-\cM_{ba}$ and 
$\cM_{\a\b}= \cM_{\b\a}$ the Lorentz generators with vector and spinor indices respectively
(see section 2 for the explicit relation between them).
These (anti-)commutation relations define the geometry of (1,1) AdS superspace.
In the case of Type II AdS supergravity, the on-shell geometry is described by covariant derivatives 
\bea
\bfD_A = (\bfD_a , \bfD_\a, \bar \bfD^\a)= {E}_A{}^M \pa_M +\hf {\O}_A{}^{cd}\cM_{cd}
+ \ri \,{ \F}_A \cJ
\label{1.4}
\eea 
obeying the following algebra: 
\bsubeq \label{20AdSsuperspace}
\bea
\{\bfD_\a,\bfD_\b\}
&=&
\{\bfDB_\a,\bfDB_\b\}
=0
~,~~~
\{\bfD_\a,\bfDB_\b\}
=
-2\ri\bfD_{\a\b}
-\ri \r \ve_{\a\b} \cJ
+\ri\r \cM_{\a\b} ~, 
\label{AdS_(2,0)_algebra_1}
\\
{[}\bfD_{\a\b},\bfD_\g{]}
&=&
-\hf \r \ve_{\g(\a} \bfD_{\b)}~,\qquad
{[}\bfD_{\a\b},\bfDB_\g{]}
=
-\hf \r \ve_{\g(\a} \bfDB_{\b)}
~,
\label{AdS_(2,0)_algebra_3/2}
\\
{[}\bfD_a,\bfD_b{]} &=& - \frac{1}{4} \r^2 \cM_{ab}~.
\label{AdS_(2,0)_algebra_2}
\eea
\esubeq
Here the constant real parameter $\r$ determines the scale of the cosmological constant,
 and $\cJ$ is the generator of the $R$-symmetry group
${\rm U(1)}_R$.
The (anti-)commutation relations (\ref{20AdSsuperspace})
define the geometry of (2,0) AdS superspace.

Comparing the relations (\ref{11AdSsuperspace}) and (\ref{20AdSsuperspace}) shows 
that  the two superspace geometries  are inequivalent, although the bosonic
bodies of the two superspaces can be shown to be identical and coincide with the ordinary AdS space. 
This indicates that properties of supersymmetric field theory in the (1,1) AdS superspace 
may considerably differ from those in the (2,0) case. 
In four dimensions, nontrivial information about supersymmetric theories defined on 
maximally symmetric superspaces   is encoded in the structure of 
consistent supercurrent multiplets associated with these superspaces
\cite{BK_AdS_supercurrent,BKsigma}. Indeed, it has been shown that 
4D $\cN=1$ rigid supersymmetric theories in AdS differ significantly from 
their counterparts defined in Minkowski space \cite{AJKL,FS,BKsigma}, and so do the 
corresponding supercurrent multiplets \cite{BK_AdS_supercurrent,BKsigma}.
This motivates us to study consistent supercurrents in the (1,1) and (2,0) AdS superspaces.

The supercurrent   \cite{FZ} is a supermultiplet  containing the energy-momentum tensor 
and the supersymmetry current(s) as well as  some other bosonic and fermionic operators.
The supercurrent naturally originates as the source of supergravity  \cite{OS,FZ2,Siegel}, 
and this realization gives a powerful practical tool to compute this multiplet for 
a given supersymmetric field theory in Minkowski space (see \cite{GGRS,Ideas} for reviews). 
Specifically, if the theory under consideration can be coupled to an off-shell supergravity background, 
then its supercurrent and associated trace multiplet coincide with 
(covariantized) variational  derivatives of the action with respect to the supergravity prepotentials 
evaluated at the background configuration corresponding to Minkowski superspace.
Since there exist several off-shell formulations for 4D $\cN=1$ supergravity \cite{old,new,non-min},
 there appear several consistent supercurrent multiplets, 
 studied e.g.~in \cite{CPS,GGS,GGRS}, of which the Ferrara-Zumino 
multiplet \cite{FZ} is usually considered to be universal.
Another useful scheme to compute supercurrents is 
the superfield Noether procedure \cite{Osborn,MSW} (which can in fact be derived from 
the off-shell supergravity  techniques presented in \cite{GGRS,Ideas}).

Recently, there has been much interest in consistent $\cN=1$ supercurrents in four dimensions
\cite{KS-FI}--\cite{DS} inspired by two papers of Komargodski and Seiberg 
\cite{KS-FI,KS2}.\footnote{General $\cN=2$ supercurrent multiplets 
in Minkowski  and AdS space were constructed 
in \cite{BK_supercurrent} and \cite{BK_AdS_supercurrent} respectively.}
These authors noticed the existence of  
certain rigid supersymmetric  theories 
for which the Ferrara-Zumino (FZ) multiplet is not well defined. 
Such theories include (i) models with a  Fayet-Iliopoulos term; and  
(ii)  supersymmetric nonlinear $\s$-models with non-exact K\"ahler forms. 
In the case (i), the appropriate supercurrent was shown in  \cite{DT,K-FI}
to be the so-called $\cR$-multiplet which is associated with  the new minimal formulation 
of $\cN=1$ supergravity \cite{new}. 
To furnish the case (ii) with a consistent supercurrent,
Ref.~\cite{KS2} put forward the so-called $\cS$-multiplet which incorporates both FZ and $\cR$ multiplets
as special limits. Although the $\cS$-multiplet can be embedded in an even more general supercurrent 
\cite{K-FI,K-Noether} of natural supergravity origin, 
it has recently been argued by Dumitrescu and Seiberg \cite{DS}
that the $\cS$-multiplet is the most general supercurrent modulo a well defined  improvement 
transformation. These authors have also derived 
a 3D $\cN=2$ super-Poincar\'e extension of the $\cS$-multiplet.
In spite of the fact  that the $\cS$-multiplet is fundamental in Poincar\'e supersymmetry, 
it does not have a natural extension to the AdS case 
in four dimensions \cite{BK_AdS_supercurrent,BKsigma}. It is also to be  expected 
that special care is required to construct consistent supercurrents for theories possessing 
the (1,1) and (2,0) AdS supersymmetry types in three dimensions. 
This problem is addressed in the present paper.

This paper is organized as follows. In section 2 we review and elaborate on the superspace 
geometry 
of  $\cN=2$ conformal supergravity presented in \cite{KLT-M}.
In sections 3 to 5 we present three superfield formulations for $\cN=2$ supergravity 
that allow for well defined  cosmological terms and supersymmetric AdS solutions.
In section 6 we describe the realizations of (1,1) and (2,0) AdS superspaces as conformally 
flat supergeometries. Section 7 presents four off-shell formulations for linearized $\cN=2$ supergravity 
in Minkowski space. Using the explicit structure of the linearized supergravity actions, 
in section 8 we construct consistent supercurrent multiplets in Minkowski space and study their properties. 
Section 9 is devoted to rigid supersymmetric theories in (1,1) AdS superspace, 
and section 10 gives a similar analysis in the (2,0) case. 
Concluding comments  are given in section 11.
The main body of the paper is  accompanied by an appendix in which we review 
the structure of 4D $\cN = 1$ supercurrents in Minkowski space.

\section{Geometry of  N = 2 conformal supergravity}
\setcounter{equation}{0}
\label{SCG}

In our recent work \cite{KLT-M}
the superspace geometry of three-dimensional ${\cal N}$-extended conformal supergravity was 
developed.  
In this section we review the formulation for $\cN=2$ conformal supergravity.

Consider a curved 3D $\cN=2$ superspace  $\cM^{3|4}$ parametrized by
local bosonic ($x$) and fermionic ($\q, \bar \q$)
coordinates  $z^{{M}}=(x^{m},\q^{\mu},{\bar \q}_{{\mu}})$,
where $m=0,1,2$, $\mu=1,2$.
The Grassmann variables $\q^{\mu} $ and $\teb_{{\mu}}$
are related to each other by complex conjugation:
$\overline{\q^{\mu}}=\teb^{{\mu}}$.
The structure group is chosen to be ${\rm SL}(2,{\mathbb{R}})\times {\rm U(1)}_R$
and the covariant derivatives
$\cD_{{A}} =(\cD_{{a}}, \cD_{{\a}},\cDB^\a)$
have the form
\bea
\cD_{{A}}&=&E_{{A}}
+\O_{{A}}
+\ri \,\F_{{A}}\cJ~.
\label{CovDev}
\eea
Here $E_{{A}}=E_{{A}}{}^{{M}}(z) \pa/\pa z^{{M}}$
is the supervielbein, 
\bea
\O_A=\hf\O_{A}{}^{bc}\cM_{bc}
=\hf\O_{A}{}^{\b\g}\cM_{\b\g}~,
\label{Lorentzconnection}
\eea
is the Lorentz connection,
and $\F_{A}$ is the ${\rm U(1)}_R$-connection.
The Lorentz generators with two  vector indices ($\cM_{ab}=-\cM_{ba}$),
one vector index ($\cM_a$) and two  spinor indices
($\cM_{\a\b}=\cM_{\b\a}$) are related to each other as follows:
$$
\cM_{a}=\hf\ve_{abc}\cM^{bc}~,~~~
\cM_{ab}=-\ve_{abc}\cM^c~,~~~
\cM_{\a\b}=(\g^a)_{\a\b}\cM_{a}~,~~~
\cM_{a}=-\hf(\g_a)^{\a\b}\cM_{\a\b}~.
$$
Here $\ve_{abc}$ ($\ve_{012}=-1$)
is the Levi-Civita tensor and $(\g_a)_{\a\b}$ are the symmetric and real
gamma-matrices defined in subsection \ref{subsection7.1}.
The generators of SL(2,$\mathbb{R})\times {\rm U(1)}_R$
act on the covariant derivatives as follows:\footnote{We refer the reader to \cite{KLT-M} for more
details on our conventions; see also subsection \ref{subsection7.1} of this paper.
 Note that  the (anti)symmetrization of $n$ indices is defined to include a factor of $(n!)^{-1}$.}
\bea
&{[}\cJ,\cD_{\a}{]}
=\cD_{\a}~,
\qquad
{[}\cJ,\cDB^{\a}{]}
=-\cDB^\a~,
\qquad
{[}\cJ,\cD_a{]}=0~,
\non \\
&{[}\cM_{\a\b},\cD_{\g}{]}
=\ve_{\g(\a}\cD_{\b)}~,\qquad
{[}\cM_{\a\b},\cDB_{\g}{]}=\ve_{\g(\a}\cDB_{\b)}~,
~~~
{[}\cM_{ab},\cD_c{]}=2\eta_{c[a}\cD_{b]}~.
\label{generators}
\eea

The supergravity gauge group is generated by local transformations
of the form
\be
\d_K \cD_{{A}} = [K  , \cD_{{A}}]~,
\qquad K = K^{{C}}(z) \cD_{{C}} +\hf K^{ c d }(z) \cM_{c d}
+\ri \, \t (z) \cJ  ~,
\label{tau}
\ee
with the gauge parameters
obeying natural reality conditions, but otherwise  arbitrary.
Given a tensor superfield $U(z)$, with its indices suppressed,
it transforms as follows:
\bea
\d_K U = K\, U~.
\label{tensor-K}
\eea

The  covariant derivatives obey (anti-)commutation relations of the form
\bea
{[}\cD_{{A}},\cD_{{B}}\}&=&
T_{ {A}{B} }{}^{{C}}\cD_{{C}}
+\hf R_{{A} {B}}{}^{{cd}}\cM_{{cd}}
+\ri \,R_{ {A} {B}}\cJ
~,
\label{algebra}
\eea
where $T_{{A} {B} }{}^{{C}}$ is the torsion,
and $R_{ {A} {B}}{}^{{cd}}$  and  $R_{{A} {B}}$
constitute the curvature.
According to the analysis given in \cite{KLT-M}, the conventional constraints \cite{HIPT}
and the Bianchi identities
lead to the spinor-spinor anti-commutation relations\footnote{For convenience, 
in the present paper the torsion superfield $\cS$ of \cite{KLT-M} 
 has been replaced by ${\mathbb S}=4\cS$.}
\bsubeq
\bea
\{\cD_\a,\cD_\b\}
&=&
-4\bar{R}\cM_{\a\b}
~,~~~~~~
\{\cDB_\a,\cDB_\b\}
=
4{R}\cM_{\a\b}~,
\label{N=2-alg-1}
\\
\{\cD_\a,\cDB_\b\}
&=&
-2\ri\cD_{\a\b}
-2\cC_{\a\b}\cJ
-\ri\ve_{\a\b} {\mathbb S} \cJ
+\ri {\mathbb S} \cM_{\a\b}
-2\ve_{\a\b}\cC^{\g\d}\cM_{\g\d} ~.
\label{dim-1-algebra}
\eea
\esubeq
The vector-spinor commutation relations are $(\cD_{\a\b}=(\g^a)_{\a\b}\cD_a)$:
\bea
{[}\cD_{\a\b},\cD_\g{]}
&=&
-\ri\ve_{\g(\a}\cC_{\b)\d}\cD^{\d}
+\ri\cC_{\g(\a}\cD_{\b)}
-\hf\ve_{\g(\a} {\mathbb  S}\cD_{\b)}
-2\ri\ve_{\g(\a}\bar{R}\cDB_{\b)}
\non\\
&&
+2\ve_{\g(\a}C_{\b)\d\r}\cM^{\d\r}
-\frac{4}{3}\Big(
\hf\cD_{(\a} {\mathbb S}
+\ri\cDB_{(\a}\bar{R}
\Big)\cM_{\b)\g}
+\frac{1}{3}\Big(
\hf \cD_{\g} {\mathbb S}
+\ri\cDB_{\g}\bar{R}
\Big)\cM_{\a\b}
\non\\
&&
+\Big(
C_{\a\b\g}
+\frac{1}{3}\ve_{\g(\a}\big(
2\cD_{\b)} {\mathbb S}
+\ri\cDB_{\b)}\bar{R}
\big)
\Big)\cJ
~.
\label{dim-3/2-algebra}
\eea
Finally, the commutator of two vector covariant derivatives
turns out to be\footnote{Note that the complete algebra of covariant derivatives is presented here for the first time.
Eq. (\ref{dim-2-algebra}) was not given in \cite{KLT-M}.}
\bea
{[}\cD_a,\cD_b{]}
&=&
\hf\ve_{abc}(\g^c)^{\a\b}\ve^{\g\d}\Big(
-\ri\bar{C}_{\a\b\d}
+\frac{\ri}{3}\ve_{\d(\a}\cDB_{\b)} {\mathbb S}
+\frac{2}{3}\ve_{\d(\a} \cD_{\b)} R
\Big)\cD_\g
\non\\
&&
+\hf\ve_{abc}(\g^c)^{\a\b}\ve^{\g\d}
\Big(
-\ri C_{\a\b\d}
+\frac{\ri}{3}\ve_{\d(\a}  \cD_{\b)}{\mathbb S}
-\frac{2}{3}\ve_{\d(\a}\cDB_{\b)}\bar{R}
\Big)\cDB_\g
\non\\
&&
-\ve_{abd}\Big{[}\,
\d^d_c\Big(
\frac{1}{6}\big(\cD^2 R+\cDB^2 \bar{R}\big)
+\frac{\ri}{3} \cD^\a\cDB_{\a}{\mathbb S}
-4\bar{R}R
-\frac{1}{4}{\mathbb S}^2
\Big)
\non\\
&&~~~~~~~~~
+\frac{\ri}{4}(\g^d)^{\a\b}(\g_c)^{\g\d}
\big(\cD_{(\a}\bar{C}_{\b\g\d)}
+\cDB_{(\a} C_{\b\g\d)}\big)
-4\cC^{d}\cC_c
\,\Big{]}\cM^c
\non\\
&&
+\frac{\ri}{8}\ve_{abc}(\g^c)^{\a\b}
\Big(
\cD^\g\bar{C}_{\g\a\b}-\cDB^\g{C}_{\g\a\b}
+ \frac{1}{3}
[\cD_\a,\cDB_{\b}]{\mathbb S}
\Big) \cJ
~.~~~
\label{dim-2-algebra}
\eea
The algebra is parametrized by three dimension-1 torsion superfields:
a real scalar $\mathbb S$, a
complex scalar $R$ and its conjugate $\bar{R}$,  and a real vector $\cC_a$
($\cC_{\a\b}:=(\g^a)_{\a\b}\cC_a$).
The superfields $\mathbb S$ and $\cC_a$  are neutral under the group ${\rm U(1)}_R$, while 
the ${\rm U(1)}_R$ charge of $R$ is $-2$, 
 $\cJ R=-2R$ and $\cJ \bar{R}=2\bar{R}$.
The torsion superfields obey differential constraints implied by the Bianchi identities.
At dimension-3/2 these are
\bea
\cDB_\a R=0~,~~~~~~
\cD_{\a}\cC_{\b\g}
=
\ri C_{\a\b\g}
+\frac{\ri }{ 3}\ve_{\a(\b}\Big(
\ri \cDB_{\g)}\bar{R}
-\cD_{\g)}{\mathbb S}
\Big)
~,
\label{N2SG-dim-3/2-constr}
\eea
together with their complex conjugates.
These equations  and their higher-dimension descendants
are sufficient to solve the complete set of Bianchi identities.
One dimension-2 descendant equation which is 
important for our subsequent analysis is 
\bea
(\cD^2
-4\bar{R})\mathbb S
=(\cDB^2
-4R)\mathbb S
&=&0
~.
\label{4.16--}
\eea
This means that the torsion $\mathbb S$ is a real covariantly linear superfield.

The rule for integration by parts  in superspace is as follows:
given a vector superfield $V=V^A E_A$,  it holds that 
\bea
\int\rd^3x\,\rd^4\q
\,  E \,(-1)^{\ve_A}  \cD_AV^A =0
~,\qquad
E^{-1}= {\rm Ber}(E_A{}^M) ~.
\label{2.12IbP}
\eea
Given a real scalar superfield $\cL$, the following chiral reduction rule also holds
\bea
\int\rd^3x \rd^4\q
\, E \,\cL = \int\rd^3x \rd^4\q
\, \frac{E}{R} \, {\bar \D}\cL
=\int\rd^3x\rd^2\q\, \cE \,\bar{\D}\cL
~,
\label{N=2Ac-2}
\eea
where $\cE$ denotes the chiral density, ${\bar \cD}_\a \cE=0$, and $\bar \D$ the chiral
projection operator 
\bea
\bar{\D}:=-\frac{1}{4}(\cDB^2-4R)~.
\label{N=2chiralPsi}
\eea

We conclude by remarking that the algebra (\ref{N=2-alg-1})--(\ref{dim-2-algebra})
and the Bianchi identities (\ref{N2SG-dim-3/2-constr})
are invariant under super-Weyl transformations generated by a real unconstrained
superfield $\s$.
The invariance under super-Weyl transformations ensures
that the geometry under consideration describes conformal supergravity.
The super-Weyl transformation of the covariant derivatives is\footnote{The infinitesimal
super-Weyl transformations, that we will denote with $\d_\s$,
 were given in our previous paper \cite{KLT-M}. Here the full nonlinear result
 is presented for the first time.}
\bsubeq\label{sW-general}
\bea
\cD'{}_\a&=&\re^{\hf\s}\Big(\cD_\a+(\cD^{\g}\s)\cM_{\g\a}-(\cD_{\a }\s)\cJ\Big)~,
\\
\cDB'{}_{\a}&=&\re^{\hf\s}\Big(\cDB_{\a}+(\cDB^{\g}\s){\cM}_{\g\a}
+(\cDB_{\a}\s)\cJ\Big)~,
\\
\cD'{}_{a}
&=&\re^{\s}\Big(
\cD_{a}
-\frac{\ri}{2}(\g_a)^{\g\d}(\cD_{(\g}\s)\cDB_{\d)}
-\frac{\ri}{2}(\g_a)^{\g\d}(\cDB_{(\g}\s)\cD_{\d)}
+\ve_{abc}(\cD^b\s)\cM^c
\non\\
&&
+\frac{\ri}{2}(\cD_{\g}\s)(\cDB^{\g}\s)\cM_{a}
-\frac{\ri}{8}(\g_a)^{\g\d}({[}\cD_{\g},\cDB_{\d}{]}\s)\cJ
-\frac{3\ri}{4}(\g_a)^{\g\d}(\cD_{\g}\s)(\cDB_{\d}\s)\cJ
\Big)~.~~~~~~
\eea
\esubeq
To ensure that the algebra (\ref{N=2-alg-1})--(\ref{dim-2-algebra}) is invariant,
the dimension-1 torsion components have to transform as
\bsubeq\label{sW-tors}
\bea
{\mathbb S}'&=&\re^{\s}\Big(
{\mathbb S}
+\ri(\cD^\g\cDB_{\g}\s)
\Big)
~,
\label{sW-S-1}
\\
\cC'_{a}&=&\re^{\s}\Big(
\cC_{a}
+\frac{1}{8}(\g_a)^{\g\d}([\cD_{\g},\cDB_{\d}]\s)
+\frac{1}{4}(\g_a)^{\g\d}(\cD_{\g}\s)(\cDB_{\d}\s)
\Big)
~,~~~~~~
\label{sW-C-1}
\\
R'&=&
\re^{\s}\Big(
R
+\frac{1}{4}(\cDB^2\s)
-\frac{1}{4}(\cDB_\g\s)(\cDB^{\g}\s)
\Big)
~,
\label{sW-R-1}
\\
\bar{R}'&=&
\re^{\s}\Big(
\bar{R}
+\frac{1}{4}(\cD^2\s)
-\frac{1}{4}(\cD^\g\s)(\cD_{\g}\s)
\Big)
~.
\label{sW-Rb-1}
\eea
\esubeq
For later use, it is useful to rewrite the transformations of the dimension-1 torsion superfields in the 
following equivalent form
\bsubeq\label{sW-tor2}
\bea
{\mathbb S}'&=&
\Big(\re^{\s}{\mathbb S}
+\ri(\cD^\g\cDB_{\g}\re^{\s})
-\ri\re^{-\s}(\cD^\g\re^{\s})(\cDB_\g\re^{\s})\Big)
~,
\\
\cC'_{a}&=&
\Big(
\cC_{a}
+\frac{1}{8}(\g_a)^{\g\d}{[}\cD_{\g},\cDB_{\d}{]}
\Big)\re^{\s}
~,~~~~~~
\\
R'&=&
-\frac{1}{4}
\re^{2\s}\Big(\cDB^2-4R\Big)\re^{-\s}
~,
~~~~~~
\bar{R}'=
-\frac{1}{4}\re^{2\s}
\Big(\cD^2-4\bar{R}\Big)\re^{-\s}
~.
\label{sW-R-2}
\eea
\esubeq

\section{Type I minimal supergravity}
\setcounter{equation}{0}

This supergravity theory is a 3D analogue of the old minimal formulation for 
4D $\cN=1$ supergravity \cite{old} (see \cite{GGRS,Ideas,WB} for reviews).
The corresponding conformal compensators are 
a covariantly chiral scalar $\F$ and  its conjugate $\bar \F$, $\bar \cD_\a \F=0$.
It is always possible to choose the chiral compensator $\F$ to have  super-Weyl weight 1/2, 
\bea
\F'=\re^{\hf \s}\F~.
\label{3.1}
\eea
This implies that its ${\rm U(1)}_R$ charge must be  $-1/2$, in accordance with the analysis in \cite{KLT-M},
\bea
\cJ \F = -\hf \F~.
\label{3.2}
\eea

The freedom to perform the super-Weyl and local 
${\rm U(1)}_R$ transformations can be used to impose the gauge
\bea
\F=1
~.
\label{3.3}
\eea
Such a gauge fixing is accompanied by  the consistency conditions \cite{KLT-M}
\bea
0=\cDB_\a\F=-\frac{\ri}{2}\F_\a~,\qquad 
0=\{\cD_\a,\cDB_\b\}\F
=
-\F_{\a\b}
+\cC_{\a\b}
-\frac{\ri}{2} \ve_{\a\b}\mathbb S
~,
\eea
and therefore
\bea
\F_\a= \mathbb S=0~,\qquad
\F_{\a\b}= \cC_{\a\b}~.
\label{3.5}
\eea

The gauge conditions (\ref{3.5}) are invariant under a combined set of super-Weyl
and U(1)$_R$ transformations. The condition ${\mathbb S}=0$ is preserved 
if the real superfield $\s$ satisfies
\bea
\ri\cD^\g\cDB_\g\s=0
~~~
\Longleftrightarrow
~~~
\s=\l+\lb~,~~~
\cDB_\a\l
=0~,~~~
\eea
with the parameter $\l$ being an arbitrary chiral superfield.
The resulting residual super-Weyl and U(1)$_R$ transformations of the Type-I geometry 
turn out to be
\bsubeq \label{sW+U(1)}
\bea
\cD'{}_\a&=&
\re^{\frac{1}{2}(3\bar{\l}-\l)}
\Big(\cD_\a+(\cD^{\g}\l)\cM_{\g\a}\Big)~,
\\
\cDB'{}_\a&=&
\re^{\frac{1}{2}(3\l-\bar{\l})}
\Big(\cDB_\a+(\cDB^{\g}\bar{\l})\cM_{\g\a}\Big)~,
\\
\cD'{}_{a}&=&\re^{\l+\bar{\l}}\Big(
\cD_{a}
-\frac{\ri}{2}(\g_a)^{\a\b}(\cD_{\a}\l)\cDB_{\b}
-\frac{\ri}{2}(\g_a)^{\a\b}(\cDB_{\a}\bar{\l})\cD_{\b}
\non\\
&&~~~~~~~
+\ve_{abc}\big(\cD^b(\l+\bar{\l})\big)\cM^c
+\frac{\ri}{2}(\cD_{\g}\l)(\cDB^{\g}\bar{\l})\cM_{a}
\Big)
~.
\eea
\esubeq
The dimension-1 torsion superfields transform according to the following equations
\bsubeq \label{sW+U(1)-torsion}
\bea
\cC'_{a}&=&\re^{\l+\bar{\l}}\Big(
\cC_{a}
-\frac{\ri}{2}\big(\cD_{a}(\l-\bar{\l})\big)
+\frac{1}{4}(\g_a)^{\a\b}(\cD_{\a}\l)(\cDB_{\b}\bar{\l})
\Big)
~,
\label{sW-C-Type-I}
\\
R'&=&
\re^{3\l-\bar{\l}}\Big(
R
+\frac{1}{4}(\cDB^2\bar{\l})
-\frac{1}{4}(\cDB_\g\bar{\l})(\cDB^{\g}\bar{\l})
\Big)
=
-\frac{1}{4}\re^{3\l}\Big(
\cDB^2
-4R
\Big)
\re^{-\bar{\l}}
~,
\label{sW-R-Type-I}
\\
\bar{R}'&=&
\re^{3\bar{\l}-\l}\Big(
\bar{R}
+\frac{1}{4}(\cD^2{\l})
-\frac{1}{4}(\cD^\g{\l})(\cD_{\g}{\l})
\Big)
=
-\frac{1}{4}\re^{3\bar{\l}}\Big(
\cD^2
-4\bar{R}
\Big)
\re^{-\l}
~.
\label{sW-Rb-Type-I}
\eea
\esubeq

\subsection{Supergravity without a cosmological term}

The supergravity action is 
\bea
S=-4\int {\rm d}^3x {\rm d}^4 \q 
\,E\,\bar \F \F~.
\label{3.6}
\eea
The equation of motion for $\F$ is 
\bea
(\cDB^2-4R)\bar \F= 0~.
\eea
In the gauge (\ref{3.3}) it reduces to 
\bea
R=0~.
\label{3.8}
\eea
Modulo purely gauge degrees of freedom,
the complete set of unconstrained prepotentials for the supergravity formulation under consideration
comprises $\F$, $\bar \F$ and a gravitational superfield $H^{\a\b} = H^{\b \a} = \overline{H^{\a\b}}$. 
In the gauge (\ref{3.3}) the equation of motion for $H^{\a\b}$ can be shown to be  
\bea
\cC_{\a\b}=0~.
\label{3.9}
\eea
The equations (\ref{3.5}), (\ref{3.8}) and (\ref{3.9}) tell us that the on-shell superspace geometry is 
locally flat.
Denoting the on-shell covariant derivatives by $D_A =(D_a, D_\a ,\bar D^\a)$, their algebra is 
\bsubeq \label{3.12}
\bea
\{D_\a,D_\b\}
&=&0
~,\qquad
\{\bar D_\a,\bar D_\b\}
=0~,
\qquad
\{D_\a,\bar D_\b\}
=
-2\ri D_{\a\b}~,\\
\big[D_a , D_\b \big] &=&0~,  \qquad \big[D_a, \bar D_\b \big]=0~, \qquad
\big[D_a,  D_b \big]=0~.
\eea
\esubeq

\subsection{Supergravity with a cosmological term}
\label{Type-I-cosmological}

The supergravity action is 
\bea
S=-4\int {\rm d}^3x {\rm d}^4 \q 
\,E\,\bar \F \F
+{\m}\int {\rm d}^3x {\rm d}^2 \q\,\cE\, \F^4
+{\bar \m}\int {\rm d}^3x  {\rm d}^2 {\bar \q}\,\bar \cE\,\bar \F^4~.
\label{3.11}
\eea
The equation of motion for $\F$ is 
\bea
\frac{1}{4}(\cDB^2-4R)\bar \F + \m \F^3= 0~.
\eea
In the gauge (\ref{3.3}) it reduces to 
\bea
R=\m =\text{const}~.
\eea
The other supergravity equations  (\ref{3.5}) and  (\ref{3.9}) 
do not change. Therefore, this theory describes AdS supergravity.
Any solution of the theory is locally an AdS superspace. In addition to
a unique maximally symmetric solution (AdS superspace), there also exist supersymmetric versions
\cite{IT} of the BTZ black hole \cite{BTZ1}.\footnote{The BTZ black hole is obtained as a discrete quotient 
of the 3D AdS space \cite{BTZ2}.  A similar realization exists for its supersymmetric extensions.}
The supersymmetry properties of the black holes in three dimensions were investigated in \cite{CH}.
Here we will only be interested in the AdS superspace.
Let  $\de_A=(\de_a,\de_\a,\deb^\a)$ be the resulting on-shell covariant derivatives, eq.~(\ref{1.2}),
obeying the (anti)commutation relations
(\ref{AdS_(1,1)_algebra_1})--(\ref{AdS_(1,1)_algebra_2}).
They describe,  according to the classification given in  \cite{AT}, 
the (1,1) AdS superspace.

There is an alternative realization of the above supergravity formulation, 
in the spirit of \cite{KMc,BM}. It is based on representing the chiral compensator  $\F$
as a composite field, 
\bea
\F^4 = -{1 \over 4} ({\bar \cD}^2 -4R) \,P~,
\qquad {\bar P} =P~, 
\eea
where $P$ is a  real unconstrained scalar with the super-Weyl transformation 
\bea
P ~ \to ~  {\rm e}^{\s  } \,P~.
\label{s-weyl-P}
\eea
The supergravity action tuns into
\bea
S=-4\int {\rm d}^3x {\rm d}^4 \q \,E\,\bar \F \F
+ ({\m} + \bar \m) \int {\rm d}^3x {\rm d}^4 \q\,E\, P~.
\label{3.19P}
\eea
The cosmological term looks like a Fayet-Iliopoulos term.
However this interpretation is somewhat misleading since the action (\ref{3.19P})
is invariant under gauge transformations 
\bea
\d P = \mathbb H~, 
 \qquad 
(\cDB^2-4R)\mathbb H= (\cD^2-4 \bar R)\mathbb H=0
\eea
which do not describe a vector multiplet, but rather a scalar multiplet.

\subsection{Matter-coupled supergravity} 
\label{subsection33}

Mater couplings in Type I supergravity are analogous to those in the old minimal formulation for
4D $\cN=1$ supergravity, see e.g. \cite{WB} for a review.
As an example,  
we only consider a general locally supersymmetric nonlinear $\s$-model
\begin{align}
S = -4 \int {\rm d}^3x {\rm d}^4 \q 
\,E\,
  \bar \F \F \,{\rm e}^{-K /4}
	+ \int {\rm d}^3x {\rm d}^2 \q \,\cE\,
	\F^4 W
	+ \int {\rm d}^3x  {\rm d}^2 {\bar \q}\,\bar \cE\, \bar\F^4 \bar W~.
	\label{3.15}
\end{align}
Here the K\"ahler potential,  $K= K(\vf^I, \bar \vf^{\bar J})$, is a real function 
of the covariantly chiral superfields $\vf^I$ and their conjugates
$\bar \vf^{\bar I}$, obeying $\bar \cD_\a \vf^I =0$.
The superpotential,  $W = W(\vf^I)$,  is a holomorphic function of $\vf^I$ alone.
The matter superfields $\vf^I$ and $\bar \vf^{\bar J}$ are chosen to be inert under the super-Weyl 
and local ${\rm U(1)}_R$ 
transformations. This guarantees the super-Weyl invariance of the action.

The action (\ref{3.1}) is invariant under K\"ahler transformations,
\begin{align}
K   \rightarrow K + F + \bar F, \qquad W \rightarrow {\rm e}^{-F} \,W, \qquad
\F \rightarrow {\rm e}^{F/4}\,\F~,
\label{3.16}
\end{align}
with  $F(\vf^I)$ an arbitrary holomorphic function.

\section{Type II minimal  supergravity}
\setcounter{equation}{0}
This supergravity theory is a 3D analogue of the new minimal formulation 
for 4D $\cN=1$ supergravity \cite{new} (see \cite{Ideas,GGRS} for reviews).
Its conformal compensator is a vector multiplet  described by  a real scalar superfield
$G$ which is defined modulo arbitrary gauge transformations of the form:
\be 
\d G = \l + \bar \l~, \qquad  \cJ \l =0~, \quad {\bar \cD}_\a \l =0~.
\ee
The gauge field is inert under the super-Weyl transformations,
\bea
 G' =G~.
\label{N=2sWrealLinear0}
\eea
Associated with $G$ is the gauge-invariant field strength 
\bea
\mathbb G= \ri {\bar \cD}^\a \cD_\a G = \bar {\mathbb G}
\label{G-prep}
\eea
which is covariantly linear,
\bea
(\cD^2 - 4 \bar R ) \mathbb G =  
(\cDB^2-4R)\mathbb G= 0~,
\label{N=2realLinear}
\eea
and is required to be nowhere vanishing, $\mathbb G \neq 0$.
The expression (\ref{G-prep}) is the most general solution to 
the constraint (\ref{N=2realLinear}).

In accordance with (\ref{N=2sWrealLinear0}),
the super-Weyl transformation of $\mathbb G$ is 
\bea
{\mathbb G}'=\re^{\s} \mathbb G~.
\label{N=2sWrealLinear}
\eea
Since $\mathbb G$ is nowhere vanishing, it is always possible to choose the super-Weyl gauge 
\bea
\mathbb G=1~.
\label{2.15}
\eea
This gauge condition will be often used in what follows.

As a consequence of (\ref{N=2realLinear}),
the gauge condition (\ref{2.15}) implies that 
\bea
R= \bar{R}=0~.
\label{4.13}
\eea
The curved superspace associated with the super-Weyl gauge choice  (\ref{4.13})
will be referred to as Type-II geometry.

Due to the equations (\ref{sW-R-2}), it is clear that the Type-II geometry 
is invariant under residual super-Weyl transformations generated by a real superfield $\s$
such that
\bea
\cD^2\re^{-\s}=\cDB^2\re^{-\s}=0
~.
\eea
Not surprisingly, the residual super-Weyl transformations are generated by a vector multiplet.

\subsection{Supergravity without a cosmological term}
The pure supergravity action \cite{KLT-M} is 
\bea
S=\int {\rm d}^3x {\rm d}^4 \q 
\,E\,L_{\text{Einst}} ~,
\label{TypeII_action}
\eea
where
\bea
L_{\text{Einst}} =4\Big(  \mathbb G \ln \mathbb G - G{\mathbb  S} \Big)~.
\label{4.8}
\eea
We recall that the torsion superfield ${\mathbb S}$ is covariantly real linear, eq. (\ref{4.16--}).
Its  super-Weyl transformation is given by (\ref{sW-S-1}).
Due to the relations (\ref{sW-S-1}),
(\ref{N=2sWrealLinear0}) and (\ref{N=2sWrealLinear}), 
the action (\ref{TypeII_action}) is invariant under the super-Weyl transformations.

Consider the equation of motion for $G$:
\bea
 \ri {\bar \cD}^\a \cD_\a \ln \mathbb G - \mathbb S =0~.
\eea
Let us choose the super-Weyl gauge (\ref{2.15}).
Then, the  equation of motion gives
\bea
\mathbb S=0~.
\label{4.12}
\eea
It should be remembered that the relation  (\ref{4.13}) holds
 in the same gauge.

The compensator $G$ is one of the two supergravity prepotentials.
The second prepotential is a gravitational superfield $H^{\a\b} = H^{\b \a} = \overline{H^{\a\b}}$. 
The corresponding equation of motion in the gauge (\ref{2.15}) is 
\bea
\cC_{\a\b}=0~.
\label{4.14}
\eea
The equations (\ref{4.13}), (\ref{4.12}) and (\ref{4.14}) tell us that the on-shell superspace geometry is locally flat.
We conclude that  this theory  describes $\cN=2$ supergravity without a cosmological term.

The supergravity formulation introduced  can equivalently be described by a
Lagrangian that   slightly differs in its form from 
(\ref{4.8}). In order to derive such a Lagrangian, a few formal 
observations should be made.
First of all, the constraint (\ref{N=2-alg-1}) implies\footnote{A complete solution to the supergravity constraints in terms of unconstrained prepotentials will be given elsewhere.}
that 
\bea
\cD_\a = E_\a +\hf \O_\a{}^{cd } \cM_{cd} - E_\a U \cJ~, 
\eea
for some complex scalar prepotential $U$ defined modulo gauge transformations
\bea
U \to U +\bar \l~, \qquad \cD_\a \bar \l=0~,
\eea
with $\l$ an arbitrary chiral scalar of zero ${\rm U(1)}_R$  charge. 
Our second observation is that the prepotential $U$ is characterized by the  super-Weyl 
and local U(1)${}_R$ transformation laws:
\begin{subequations}
\bea
\d_\s U &=& \s~, \label{4.17a} \\
\d_\t U &=& \ri \, \t~. \label{4.17b}
\eea
\end{subequations}
Finally, the third observation is that the constraint (\ref{dim-1-algebra}) 
leads to the following relation
\bea
{\mathbb S} = \ri  \cD^\a\bar \cD_\a S~, \qquad
S=\hf (U +\bar U)~.
\eea
Now, integration by parts can be used to show  that the Lagrangian (\ref{4.8}) is equivalent to 
\bea
\tilde{L}_{\text{Einst}} =4 \mathbb G \Big( \ln \mathbb G - S \Big)~.
\label{4.8-mod}
\eea

The above supergravity theory is dual to that described by the action 
(\ref{3.6}). To prove this, it suffices to consider the following first-order model:
\bea
L_{\text{first-order}} = 4 \bG \Big( \ln \bG - 1- S - \j - \bar \j \Big)~, 
\qquad \bar \cD_\a \j =0~.
\label{first-order}
\eea
Here $\bG$ is a real unconstrained superfield, and $\j $ a chiral scalar of zero ${\rm U(1)}_R$ charge.
Varying the action with respect to $\j$ gives $\bG = \mathbb G$, and then the model under consideration reduces 
to that described by the Lagrangian (\ref{4.8-mod}). On the other hand, the auxiliary superfield $\bG$ can be
integrated out using its equation of motion, $\ln \bG = S + \j +\bar \j$.
This leads to the supergravity theory (\ref{3.6}) in which 
\bea
\F:= \re^{  \hf \bar U } \re^{ \j }~.
\eea
Using the  super-Weyl 
and local ${\rm U(1)}_R$  transformation laws of $U$, eqs. (\ref{4.17a}) and (\ref{4.17b}), 
one may see that $\F$ is a covariantly chiral superfield 
characterized by the properties (\ref{3.1}) and (\ref{3.2}).

\subsection{Supergravity with a cosmological term}
\label{Type-II-cosmological}

Consider a deformed supergravity action 
\bea
S_{\text{AdS}}=\int {\rm d}^3x {\rm d}^4 \q 
\,E\,L_{\text{AdS}} ~,
\eea
where, up to a total derivative,
\bea
L_{\text{AdS}}  = 4\Big( \mathbb G \ln \mathbb G - G{\mathbb S} +
\hf\r
G \mathbb G \Big)
\simeq 4 \mathbb G\Big( \ln \mathbb G - S + \hf\r G \Big)
~,
\label{Type-II-AdS}
\eea
with $\r$ a real coupling constant. This Lagrangian differs from (\ref{4.8}), or its equivalent form 
(\ref{4.8-mod}),  
by the presence 
of a Chern-Simons term.

Now, the equation of motion for $G$ is
\bea
 \ri {\bar \cD}^\a \cD_\a \ln \mathbb G - \mathbb S +\r \mathbb G=0~.
\eea
Choosing the super-Weyl gauge (\ref{2.15}) gives 
\bea
\mathbb S=\r =\text{const}~.
\eea
The supergravity equations  of motion (\ref{4.13}) and (\ref{4.14}) do  not change.
Therefore, the theory describes AdS supergravity.
Any solution to the supergravity equations of motion is locally an AdS space-time.

The algebra of the on-shell covariant derivatives becomes 
(\ref{AdS_(2,0)_algebra_1})--(\ref{AdS_(2,0)_algebra_2}).
According to the classification given in \cite{AT}, 
such an algebra describes the (2,0) AdS supergeometry.

\subsection{Matter-coupled supergravity} 
\label{subsection43}

The pure supergravity model (\ref{4.8-mod}) can be readily  generalized to include supersymmetric chiral matter
that is neutral under the local ${\rm U(1)}_R$ group
\bea
L =4 \mathbb G \Big( \ln \mathbb G - S
+ \frac{1}{4} K(\vf^I, \bar \vf^{\bar J}) \Big)
~, \qquad \bar \cD_\a \vf^I =0~
\label{sigma-Type-II}
\eea
with $K$ the K\"ahler potential of a K\"ahler manifold.
The corresponding action is 
invariant under K\"ahler transformations,
\begin{align}
K   \rightarrow K + F + \bar F~,
\label{Type-II-kahler-inv}
\end{align}
with  $F(\vf^I)$ an arbitrary holomorphic function.
The model (\ref{sigma-Type-II}) proves to be dual to (\ref{3.15}) with $W(\vf)=0$.
This duality can be demonstrated by making use of a natural generalization of the first-order
Lagrangian (\ref{first-order}).

Similarly to the new minimal $\cN=1$ supergravity in four dimensions, 
Type II minimal supergravity can be coupled to $R$-invariant $\s$-models.
Let us consider a system of self-interacting covariantly chiral superfields $\f^I$,  
where $I=1,\cdots,m$,  
with ${\rm U(1) }_R$ charges  
\bea
&
\cJ \f^I= -r_I \f^I \qquad \qquad \text{(no sum)} 
\label{charges-phi-R-system}
\eea
and hence their infinitesimal super-Weyl transformation laws are
 \bea
  \d_\s \phi^I=r_I \s\phi^I~.
 \eea
In order to have an $R$-invariant system, 
the K\"ahler potential $K(\f^I , \bar \f^{\bar J} )$
and the superpotential $W (\f^I)$ should obey the equations
\begin{subequations}
\bea
&& \sum_I r_I \f^I K_I=\sum_{\bar{I}} r_I \bar \f^{\bar{I}} K_{\bar{I}}~, 
\label{K-R-symmetric}\\
&& \sum_I r_I \f^I W_I=2W~.
\label{W-R-symmetric}
\eea
\end{subequations}
The complete supergravity-matter system is described by the action
 \bea
 S= 4 \int {\rm d}^3x {\rm d}^4 \q 
 \,E\,
\mathbb G \Big( \ln \mathbb G - S
&+& \frac{1}{4} 
{K}\big(\phi^I/{\mathbb G}^{r_I}, \bar {\phi}^{\bar J}/{\mathbb G}^{r_J}\big)
\Big)
\non \\
+  \Big{\{} \int {\rm d}^3x {\rm d}^2 \q \,\cE\,
{W}(\phi^I)
\, &+&\,{\rm c.c.}~\Big{\}}	~.
	\label{333}
 \eea
The action can be seen to be super-Weyl invariant.
In the case of a superconformal $\s$-model, such that $r_I =1/2$ and  $K(\f^I , \bar \f^{\bar J} )$
obeys the homogeneity condition 
\bea
 \sum_I  \f^I K_I= K~,
\eea
the matter sector in (\ref{333}) decouples from the  linear compensator $\mathbb G$.
 
Given a system of  Abelian vector multiplets described by gauge prepotentials $F^i$ 
and gauge invariant field strengths ${\mathbb F}^i =\ri \cD^\a \bar \cD_\a F^i$,
their coupling to supergravity can be described by an action of the form
\bea
S=\int {\rm d}^3x {\rm d}^4 \q \,E\,
{\mathbb G}\,\Big(
{\rm L}({\mathbb F}^i/{\mathbb G})
+\frac{1}{2{\mathbb G}} m_{ij}F^i{\mathbb F}^j
+ \x_{i}F^i
\Big)
~,
\label{arb-L-2}
\eea
where 
the parameters
$$
 m_{ij} =m_{ji} =\overline{m_{ij}}={\rm const}
 $$
 describe Chern-Simons couplings,  
and  $\x_i$ correspond to  Fayet-Iliopoulos terms.
If the Lagrangian $L({\mathbb F}^i)$ corresponds to a superconformal system, 
\be
{\mathbb F}^i \frac{\pa}{\pa {\mathbb F}^i}L ({\mathbb F}) =L ({\mathbb F})~,
\ee
and no Fayet-Iliopoulos term is present, $\x_i =0$, then the action (\ref{arb-L-2})
is independent of the linear compensator $\mathbb G$.

The action for Type I AdS supergravity, eq. (\ref{3.11}), is a special case of  models
(\ref{3.15}) which describe the most general coupling of conformal supergravity to 
chiral scalar multiplets. Type II AdS supergravity can also be understood as a special 
case of  models describing the most general coupling of conformal supergravity to vector multiplets.

\section{Non-minimal supergravity}
\setcounter{equation}{0}
In this section we present 3D analogues of the following 4D $\cN=1$ theories: (i) the non-minimal supergravity
 without a cosmological term \cite{non-min,SG};
and (ii) the non-minimal AdS supergravity \cite{BKdual}.

\subsection{Supergravity without a cosmological term}
\label{NM-sugra-NL}

This supergravity formulation involves the following conformal compensators:
a complex linear superfield $\S$ and its conjugate $\bar \S$. 
The superfield  $\S$ obeys the  constraint 
\bea 
(\cDB^2-4R)\S=0
\label{complex-linear}
\eea
and no reality condition. If $\S$ is chosen to transform homogeneously under the 
super-Weyl transformations, 
then its  ${\rm U(1)}_R$ charge is determined by the super-Weyl weight \cite{KLT-M} 
\bea
\d_\s \S=w\s\S \quad \Longrightarrow \quad \cJ\S=(1-w) \S
~.
\label{complex-linear2}
\eea

We derive the non-minimal supergravity action by dualizing the Type I minimal  action (\ref{3.6}).
Let us consider the first-order action 
\bea
S_{\text{first-order}}=\int {\rm d}^3x {\rm d}^4 \q 
\,E\,\Big\{ -4\bar \F \F
+\frac{2}{1-w}\big( \S \F^{2(1-w)}  +  \bar \S \bar \F^{2(1-w)}     \big) \Big\}
~,
\label{NM3}
\eea
where $\F$ is {\it complex unconstrained}, and $\S$ is complex linear.
This action is super-Weyl invariant provided $\F$ transforms as in (\ref{3.1}).
It is also invariant under local ${\rm U(1)}_R$ transformations if the 
${\rm U(1)}_R$  charge of $\F$ is chosen as in (\ref{3.2}).

The theory (\ref{NM3}) is equivalent to the  Type I minimal supergravity, eq.  (\ref{3.6}).
Indeed, varying  (\ref{NM3}) with respect to $\S$ gives $\bar \cD_\a \F =0$, and then 
(\ref{NM3}) reduces to the action (\ref{3.6}).
On the other hand, we can start from (\ref{NM3}) and integrate out the fields $\F$ and $\bar \F$.
This yields 
\bea
S_{\text{non-minimal}}= 4 \frac{w}{1-w}
\int {\rm d}^3x {\rm d}^4 \q 
\,E\,
\big(\bar \S \S\big)^{\frac{1}{2w}} ~.
\label{NM4}
\eea
This action is not defined if $w=1$. This value proves to correspond to the Type II minimal 
supergravity. It may be seen that the case of Type I  minimal supergravity corresponds
to the limit $w\to \infty$. 
The last singular point of the action (\ref{NM4}) is given by $w=0$. In this case the complex linear 
superfield is super-Weyl invariant and cannot be used as a conformal compensator.

For later use, it is worth presenting a relationship between the 3D parameter $w$ and the 
4D Siegel-Gates parameter $n$ \cite{SG}.
By identifying the U(1)${}_R$ charges of a complex linear superfield coupled to conformal 
supergravity respectively in 4D and 3D one gets the relation
\bea
\frac{4n}{3n+1}=1-w 
\label{n--w}
\eea
which is equivalent to (\ref{1.1}).

\subsection{Supergravity with a cosmological term}
\label{NM-cosmological}

The non-minimal formulation developed in the previous subsection
 is not suitable to describe AdS supergravity,
in complete analogy with the four-dimensional case \cite{GGRS}. 
In four dimensions, however, the way out has been found in \cite{BKdual}.
The same idea can successfully be applied in three dimensions.

Our point of departure will be the following super-Weyl transformation law \cite{KLT-M}
\bea
\d_\s \Big((\cDB^2-4R)\G\Big)=(1+w) \s(\cDB^2-4R)\G~,
\label{NM5}
\eea
which holds for any complex superfield $\G$ with the transformation properties
\bea
\d_\s \G= w\s\G~,\qquad
\cJ\G=(1-w)\G~.
\label{NM000}
\eea
The complex linear compensator $\S$ is an example of such a superfield.
Eq. (\ref{NM5}) 
tells us that $(\cDB^2-4R)\G$ is super-Weyl invariant if and only if $w=-1$.
In that case, we may consistently deform the linear constraint, eq.  (\ref{complex-linear}).
In what follows, we fix $w=-1$.

We introduce a new conformal compensator $\G$ which has the transformation properties
\bea
\d_\s \G= -\s\G~,\qquad
\cJ\G=2\G
\eea
and  obeys  the {\it improved} linear constraint\footnote{In global 4D $\cN=1$ supersymmetry, constraints  
of the form (\ref{NM6}) were introduced for the first time by Deo and Gates \cite{DG85}.
In the context of supergravity, such constraints have recently been used in \cite{KTyler} to generate
couplings of the Goldstino superfield to chiral  matter.}
\begin{align}
-\frac{1}{4} (\bar \cD^2 - 4 R) \Gamma = W(\vf)~,
\label{NM6}
\end{align}
with $W(\vf)$ the matter superpotential defined in  subsection \ref{subsection33}.
Using $\G$ and its conjugate $\bar \G$, we can develop a dual formulation of the theory (\ref{3.15}).
In order to achieve that, we consider the first-order action
\bea
S_{\text{first-order}} = \int {\rm d}^3x {\rm d}^4 \q 
\,E\,
\Big( -4  \bar \F \F \,{\rm e}^{-K /4}
+ \G\,\F^4 + \bar \G \,\bar\F^4 \Big) ~,
\label{NM7}
\eea
where $\F$ is {\it complex unconstrained}, and $\G$ obeys the constraint (\ref{NM6}).
Varying $S_{\text{first-order}} $ with respect to $\G$ yields $\bar \cD_\a \F =0$, 
and then the action reduces to the supergravity matter action (\ref{3.15}). 
On the other hand, we can integrate out the fields $\F$ and $\bar \F$ 
to end up with the dual model
\bea
S = -2 \int {\rm d}^3x {\rm d}^4 \q 
\,E\,
{ {\rm e}^{-K /2} }{ { (\bar \G \, \G)} }^{-1/2}~.
\label{NM8}
\eea

To describe pure AdS supergravity, we have to set $K=0$ and $W=\m$.
 Now the compensator obeys the constraint
\begin{align}
-\frac{1}{4} (\bar \cD^2 - 4 R) \Gamma = \m ={\rm const}~,
\end{align}
and the action (\ref{NM8}) turns into 
AdS supergravity 
\bea
S_{\text{AdS}} = -2 \int {\rm d}^3x {\rm d}^4 \q 
\,E\,
{ 
{ (\bar \G \, \G)} 
}^{-1/2}~.
\eea
By construction, this theory is dual to the Type I  minimal AdS supergravity, eq. (\ref{3.11}).

\section{Conformal flatness of the AdS superspaces}
\setcounter{equation}{0}

It is well known that 4D $\cN=1$ AdS superspace is conformally flat, see e.g. \cite{Ideas} 
for a pedagogical review. The same property is characteristic of the 4D $\cN=2$ 
\cite{BILS,4D-N=2-conf-flat}  and 5D $\cN=1$  \cite{5D-N=1-conf-flat} AdS superspaces.
At the same time, it was shown in \cite{BILS} that  
the conventional superspace
extensions of the coset manifolds ${\rm AdS}_2 \times S^2$,  ${\rm AdS}_3 \times S^3$ 
and ${\rm AdS_5} \times S^5$, which arise as
solutions of certain   supergravity theories in  four, six and ten dimensions, are not conformally flat.
In this section we  prove that the (1,1) and (2,0) AdS superspaces are conformally flat.
Our proof is constructive and provides explicit realizations of the  (1,1) and (2,0) AdS superspace geometries. 

\subsection{(1,1) AdS superspace}
The super-Weyl and U(1)$_R$ transformations of the Type-I  curved superspace geometry 
are given by eq. (\ref{sW+U(1)}). Our goal is to show that the covariant derivatives $\nabla_A$
of the (1,1) AdS superspace can be brought to the form
\bsubeq \label{6.1}
\bea
\de_\a&=&
\re^{\frac{1}{2}(3\bar{\l}-\l)}
\Big(D_\a+(D^{\g}\l)\cM_{\g\a}\Big)~,
\\
\deb_\a&=&
\re^{\frac{1}{2}(3\l-\bar{\l})}
\Big(\DB_\a+(\DB^{\g}\bar{\l})\cM_{\g\a}\Big)~,
\\
\de_{a}&=&\re^{\l+\bar{\l}}\Big(
\pa_{a}
-\frac{\ri}{2}(\g_a)^{\a\b}(D_{\a}\l)\DB_{\b}
-\frac{\ri}{2}(\g_a)^{\a\b}(\DB_{\a}\bar{\l})D_{\b}
\non\\
&&~~~~~~~
+\ve_{abc}  \pa^b(\l+\bar{\l}) \cM^c
+\frac{\ri}{2}(D_{\g}\l)(\DB^{\g}\bar{\l})\cM_{a}
\Big)
~,
\label{AdS-1-1-vector-dev}
\eea
\esubeq
for some chiral scalar $\l$.
Here,  $D_A =(\pa_a,D_\a,\DB^\a)$ are the flat global covariant derivatives
\bea
&&\pa_a=\frac{\pa}{\pa x^a}~,~~~
D_\a=\frac{\pa}{\pa\q^\a}+\ri\qb^\b(\g^a)_{\a\b}\pa_a
~,~~~
\DB_\a=-\frac{\pa}{\pa\qb^\a}-\ri\q^\b(\g^a)_{\a\b}\pa_a~.
\eea
They obey the (anti-)commutation relations (\ref{3.12}).

Under the super-Weyl and U(1)$_R$ transformations, the dimension-1 torsion superfields
transform according to  eq. (\ref{sW+U(1)-torsion}).
Since Minkowski superspace has no dimension-1 torsion, the chiral parameter $\l$ and its conjugate $\bar \l$
in (\ref{6.1}) must obey the following equations: 
\bsubeq
\bea
\mu&=&
-\frac{1}{4}\re^{3\l}\DB^2\re^{-\bar{\l}}
~,~~~~~~
\bar{\mu}=
-\frac{1}{4}\re^{3\bar{\l}}D^2\re^{-\l}
~,
\label{conf-flat-(1,1)-1}
\\
0&=&
\ri\pa_{\a\b}(\l-\bar{\l})
+(D_{(\a}\l) \DB_{\b)}\bar{\l} ~.
\label{conf-flat-(1,1)-2}
\eea
\esubeq
Here the complex parameter $\mu$ is the constant curvature of the 
(1,1) AdS superspace  (\ref{AdS_(1,1)_algebra_1})--(\ref{AdS_(1,1)_algebra_2}).

To find a solution of the equations (\ref{conf-flat-(1,1)-1}) and(\ref{conf-flat-(1,1)-2})
we first observe that (\ref{conf-flat-(1,1)-2}) can equivalently be rewritten as
\bea
[D_{(\a},\DB_{\b)}]\re^{\l+\lb}=0~.
\label{conf-flat-(1,1)-2-2}
\eea
We look for a  Lorentz invariant solution of this equation 
of the form
\bea
\re^{\l+\lb}
=
1+a\mu\mub x^2
+\bar{b}\mub\q^2
+b\mu\qb^2+\frac{a}{2}\mu\mub\q^2\qb^2
~,
\label{(1,1)-ansatz-1}
\eea
where 
\bea
&&x^2:=x^ax_a~,~~~
\q^2:=\q^\a\q_{\a}~,~~~
\qb^2:=\qb_\a\qb^{\a}=\overline{\q^2}
~.
\eea
The right-had side of  (\ref{(1,1)-ansatz-1})
involves  two parameters, $a$  and  $b$, which are real and complex respectively.
The superfield $\re^{\l+\lb}$ in (\ref{(1,1)-ansatz-1})
 is reminiscent of that emerging in the 4D $\cN=1$ AdS superspace geometry \cite{Ideas}.
The relation  (\ref{(1,1)-ansatz-1}) implies that 
\bea
\re^{\l}=(1+a\mu\mub\, x^2_L+2\bar{b}\mub\,\q^2)^\hf~,~~~~~~
\re^{\lb}=(1+a\mu\mub\, x^2_R+2b\mu\,\qb^2)^\hf~,
\label{(1,1)-ansatz-2}
\eea
where we have introduced the (anti)chiral vector variables
\bsubeq
\bea
&&x^a_L=x^a+\ri(\g^a)_{\a\b}\q^\a\qb^{\b}~,~~~~~~~~\qquad \quad
\DB_\a x^a_L=0
~,
\\
&&x^a_R=x^a-\ri(\g^a)_{\a\b}\q^\a\qb^{\b}
~,~~~~~~~~\qquad \quad
D_\a x^a_R=0
~.
\eea
\esubeq

Plugging (\ref{(1,1)-ansatz-2})
into equations (\ref{conf-flat-(1,1)-1}), after some 
algebra we find that (\ref{(1,1)-ansatz-2}) is indeed a solution of  (\ref{conf-flat-(1,1)-1}) provided 
$a=-1$ and  $b=-1$.
As a result,  we have constructed
an explicit conformally flat realization for the (1,1) AdS superspace.    
The covariant derivatives are given by the relations (\ref{6.1})  with
\bea
\re^{\l}=(1-\mu\mub\, x^2_L-2\mub\,\q^2)^\hf~,~~~~~~
\re^{\lb}=(1-\mu\mub\, x^2_R-2\mu\,\qb^2)^\hf~.
\eea

Using the expression for $\re^{\l+\lb}$
and the explicit form of the vector covariant derivative $\nabla_a$, 
eq. (\ref{AdS-1-1-vector-dev}), we can read off  the space-time metric 
\be
{\rm d}s^2 = {\rm d}x^a  \,{\rm d}x_a \,\big(\re^{-2(\l+\lb)}\big) \big|_{\q=0}
= \frac{{\rm d}x^a  {\rm d}x_a }{\big(1-\mu\mub x^2\big)^2}~.
\label{metric}
\ee  
This coincides  with a standard expression for the metric  of AdS${}_3$
computed using the  stereographic projection for an AdS hyperboloid.\footnote{See, e.g,
Appendix D of \cite{4D-N=2-conf-flat}
for details about the stereographic projection for ${\rm AdS}_d$.}
 As such, the  conformally flat representation is defined only locally.

\subsection{(2,0) AdS superspace}
\label{subsection6.2}

In three-dimensional $\cN=2$ supergravity,  the super-Weyl transformation of the covariant derivatives
is given by  (\ref{sW-general}).
Our goal in this subsection is to show that 
the   covariant derivatives $ {\bf D}_A$ of the (2,0) AdS superspace
can be brought to the conformally flat form:
\bsubeq
\bea
\bfD_\a&=&\re^{\hf\s}\Big(D_\a+(D^{\g}\s)\cM_{\g\a}-(D_{\a }\s)\cJ\Big)~,
\label{Dal_2-0}
\\
\bfDB_{\a}&=&\re^{\hf\s}\Big(\DB_{\a}+(\DB^{\g}\s){\cM}_{\g\a}
+(\DB_{\a}\s)\cJ\Big)~,
\label{DalB_2-0}
\\
\bfD_{a}
&=&\re^{\s}\Big(
\pa_{a}
-\frac{\ri}{2}(\g_a)^{\g\d}(D_{(\g}\s)\DB_{\d)}
-\frac{\ri}{2}(\g_a)^{\g\d}(\DB_{(\g}\s)D_{\d)}
+\ve_{abc}(\pa^b\s)\cM^c
\non\\
&&
+\frac{\ri}{2}(D_{\g}\s)(\DB^{\g}\s)\cM_{a}
-\frac{\ri}{2}(\g_a)^{\g\d}(D_{\g}\s)(\DB_{\d}\s)\cJ
\Big)
~,
\label{Da_2-0}
\eea
\esubeq
for some real scalar $\s$.
Under  the super-Weyl transformation, the dimension-1 components of the torsion 
transform according to (\ref{sW-tor2}). Since there is no dimension-1 torsion in Minkowski 
superspace, 
the super-Weyl parameter $\s$ must obey the following equations:
\bsubeq
\bea
\r&=&
\ri\re^{\s}D^\g\DB_{\g}\s
=\ri\Big(
D^\g\DB_{\g}\re^{\s}
-\re^{-\s}(D^\g\re^{\s})\DB_\g\re^{\s}\Big)~,
\label{conf-flat-(2,0)-1}
\\
0&=&
[D_{(\a},\DB_{\b)}]\re^{\s}
~,
\label{conf-flat-(2,0)-2}
\\
0&=&
\DB^2\re^{-\s}
= D^2\re^{-\s} ~.
\label{conf-flat-(2,0)-3}
\eea
\esubeq
Here  $\r$ is the parameter which appears in the (anti-)commutation relations 
(\ref{AdS_(2,0)_algebra_1})--(\ref{AdS_(2,0)_algebra_2}).
We now turn to deriving a Lorentz invariant solution  
of the  equations (\ref{conf-flat-(2,0)-1})--(\ref{conf-flat-(2,0)-3}).

It should be remarked that  the system  (\ref{conf-flat-(2,0)-1})--(\ref{conf-flat-(2,0)-3})
involves only two independent equations 
since eq.  (\ref{conf-flat-(2,0)-3})  proves to be
 a  consequence of (\ref{conf-flat-(2,0)-1}).
Indeed, eq. (\ref{conf-flat-(2,0)-3}) states that the superfield $\re^{-\s}$ is real linear,
and this automatically holds if (\ref{conf-flat-(2,0)-1}) is satisfied.
Therefore it suffices to  focus on the equations (\ref{conf-flat-(2,0)-1})
and (\ref{conf-flat-(2,0)-2}) only.

Let us start by analyzing eq. (\ref{conf-flat-(2,0)-2}).
Note that this equation has the same functional form as  (\ref{conf-flat-(1,1)-2-2})
with $\l+\lb$ replaced by $\s$.
We recall that, in searching for a solution to the system of equations 
(\ref{conf-flat-(1,1)-1}) and (\ref{conf-flat-(1,1)-2-2}),
we started with a simple ansatz (\ref{(1,1)-ansatz-1}).
That expression consists of three parts that separately satisfy (\ref{conf-flat-(2,0)-2}), 
which are:  $x^2+\hf\q^2\qb^2$,  $\q^2$ and $\bar \q^2$.
The term  proportional to a linear combination of 
$\q^2 $ and $\bar \q^2$ had to be included in (\ref{(1,1)-ansatz-1}), since $\re^{\l +\bar \l}$
should be the product of a chiral and antichiral superfields.
In the (2,0) case, however, this is not the case; in particular,  the presence of such a term would be 
inconsistent with the real linear constraint on $\re^{-\s}$. But a natural way to make an
ansatz consistent
with eq. (\ref{conf-flat-(2,0)-2})--(\ref{conf-flat-(2,0)-3}) is  to include a term 
proportional to $\ri\q^\g\qb_\g$.
These considerations lead to the ansatz
\bea
\re^{\s}
=
1+c\r^2 x^2
+\ri d\r\,\q^\g\qb_\g+\frac{c}{2}\r^2\q^2\qb^2
~,
\label{(2,0)-ansatz-1}
\eea
where $c,d$ are two constant real parameters.
Such a superfield trivially satisfies equation (\ref{conf-flat-(2,0)-2}).
After some algebra, one can prove that the function  (\ref{(2,0)-ansatz-1}) also satisfies
equation (\ref{conf-flat-(2,0)-1}) provided  the parameters $b,c$ are fixed as follows:
$c=- {1}/{16}$ and  $d=-{1}/{2}$.
We thus have constructed the  Lorentz invariant solution to 
the equations  (\ref{conf-flat-(2,0)-1})--(\ref{conf-flat-(2,0)-3}):
\bea
\re^{\s}=
1-\frac{1}{16}\r^2 x^2 
-\frac{\ri}{2}\r\,\q^\a\qb_{\a}
-\frac{1}{32}\r^2\q^2\qb^2
~.
\eea

As pointed out earlier, 
the equation (\ref{conf-flat-(2,0)-1})
implies that 
\bea
\re^{-\s}=
\frac{1}{1-\frac{1}{16}\r^2 x^2 }
+\frac{\ri\r\, \q^\a\qb_{\a}}{2\big(1-\frac{1}{16}\r^2 x^2 \big)^2}
-\frac{\r^2\q^2\qb^2\big(3+\frac{1}{16}\r^2 x^2\big)}{32\big(1-\frac{1}{16}\r^2 x^2 \big)^3}
\eea
 is real linear.  This means that $\re^{-\s}$  can be interpreted as 
 the field strength of a particular vector multiplet in flat superspace, 
\bea
{\mathbb G}_{\text{flat}} \equiv \re^{-\s}= \ri D^\a \bar D_\a G_0 ~, \qquad G_0 = \frac{1}{\r}\s~,
\label{G-flat}
\eea
such that its prepotential, $G_0$, is proportional to 
\bea
\s=
\log{\Big(1-\frac{1}{16}\r^2 x^2 \Big)}
-\frac{\ri\r\,\q^\a\qb_{\a}}{2\big(1-\frac{1}{16}\r^2 x^2 \big)}
+\frac{\r^2\q^2\qb^2\big(1+\frac{1}{16}\r^2 x^2\big)}{32\big(1-\frac{1}{16}\r^2 x^2 \big)^2}
~.
\eea
Now, we should recall the super-Weyl transformation laws of (i) the field strength $\mathbb G$ of a vector 
multiplet, eq. (\ref{N=2sWrealLinear});  and (ii)  the corresponding prepotential $G$, 
eq.  (\ref{N=2sWrealLinear0}).
Let us apply the super-Weyl transformation generated by (\ref{(2,0)-ansatz-1}), 
which takes us from Minkowski superspace to (2,0) AdS superspace, 
to the vector multiplet (\ref{G-flat}). We then end up with a vector multiplet in (2,0) AdS superspace
which is characterized  by the prepotential 
$ G_0 = \s / \r $ and the field strength 
\bea
{\mathbb G}_{\text{AdS}} = \ri \,\bfD^\a\bfDB_\a G_0 =1~.
\eea
The existence of such a frozen vector multiplet with constant field strength 
is of importance in the study of matter multiplets in 
(2,0) AdS superspace. It can be used to describe chiral scalar multiplets with a real mass generated by a
central charge.

\section{Linearized supergravity  models in Minkowski space}
\setcounter{equation}{0}

In this section we derive linearized 3D $\cN=2$ supergravity actions  by 
dimensional reduction and truncation of 4D $\cN=1$ supergravity models. 
Dimensional reduction of any off-shell 4D $\cN=1$ supergravity multiplet to three dimensions
should result in an off-shell 3D  $\cN=2$ supergravity theory coupled to a  vector/scalar  multiplet.
At the linearized level, the reduced action should be equivalent to a sum 
of decoupled supergravity and vector/scalar multiplet actions.

According to the classification of linearized off-shell actions for 4D $\cN=1$
supergravity given in \cite{GKP},  there are three minimal 
models with $12+12$ degrees of freedom and one non-minimal model (parametrized by a real 
parameter $n\neq -1/3, 0$) with $20+20$ degrees of freedom.
One can also consider reducible supergravity actions with $16+16$ degrees of freedom
obtained as a linear combination of two minimal models.

\subsection{Type II minimal supergravity}
\label{subsection7.1}

It appears that the procedure of dimensional reduction $\rm 4D \to 3D$ is simplest in the case of 
the linearized action of new minimal 4D $\cN=1$ supergravity.
This action is (see \cite{Ideas,GKP} for derivations)
\bea
S^{({\rm II})} [H_{\a\dot \a}, {\mathbb F}] &=&  \int \rd^4x\rd^4\q \,
\Big\{
-\frac{1}{16}H^{\a\bd}D^\g\DB^2 D_\g H_{\a \bd}
-\frac{1}{4}(\pa_{\a\bd}H^{\a\bd})^2 \non \\
&&\qquad \quad
+\frac{1}{16}([D_\a,\DB_\bd]  H^{\a\bd})^2
+\hf {\mathbb F}{[}D_\a,\DB_\bd{]} H^{\a\bd}
+\frac{3}{2}{\mathbb F}^2
\Big\}~.
\label{II-4D}
\eea
It is described in terms of a gravitational superfield, $H_{\a \bd} = \overline{H_{\b\ad}}$,  
and  a real linear compensator,  ${\mathbb F} = \bar{\mathbb F}$, subject 
  to the constraint $D^2{\mathbb F}={\bar D}^2 {\mathbb F} =0$.
The action is invariant under the gauge transformations
\begin{subequations}\label{5.2}
\bea
\d H_{\a \bd} &=& \bar D_{\dot\b}L_{\a}
-D_\a\bar L_{\dot\b}~,  \\
\d {\mathbb F}&=&\frac 14(D^\a\bar D^2 L_\a+\bar D_{\dot\a}D^2 \bar L^{\dot\a})~,
\eea
\end{subequations}
with $L_\a$ an unconstrained spinor parameter. 

Our goal is to dimensionally reduce the action (\ref{II-4D}) to three dimensions.
We follow \cite{KLT-M} to relate  our 3D spinor formalism  to 
the 4D sigma-matrices
\bea
(\s_{\mun } )_{\a \dot \b}:= ({\mathbbm 1}, \vec{\s} ) ~, \qquad
(\tilde{\s}_{\mun } )^{{\dot \a}  \b}:= 
\ve^{\b \g}\ve^{\ad\dd}(\s_{\mun } )_{\g \dot \d}=
({\mathbbm 1}, - \vec{\s} ) ~, \qquad {\mun }=0,1,2,3
~,
\label{4DsigmaM}
\eea
where $\vec{\s}=(\s_1,\s_2,\s_3)$ are the Pauli matrices.
By deleting the matrices with space index $\mun =2$ we obtain the 3D gamma-matrices
\begin{subequations}
\bea
(\s_{\mun } )_{\a \dot \b}\quad & \longrightarrow & \quad (\g_m )_{\a  \b} = (\g_m)_{\b\a} ~
=({\mathbbm 1}, \s_1, \s_3) ~,\\
(\tilde{\s}_{\mun } )^{\dot \a  \b}\quad & \longrightarrow & \quad (\g_m )^{\a  \b} = (\g_m)^{\b\a}
=\ve^{\a \g} \ve^{\b \d} (\g_m)_{\g \d} ~,
\eea
\end{subequations}
where the spinor indices are  raised and lowered using
the SL(2,${\mathbb R}$) invariant tensors
\bea
\ve_{\a\b}=\left(\begin{array}{cc}0~&-1\\1~&0\end{array}\right)~,\qquad
\ve^{\a\b}=\left(\begin{array}{cc}0~&1\\-1~&0\end{array}\right)~,\qquad
\ve^{\a\g}\ve_{\g\b}=\d^\a_\b
\eea
as follows:
\bea
\psi^{\a}=\ve^{\a\b}\psi_\b~, \qquad \psi_{\a}=\ve_{\a\b}\psi^\b~.
\eea
By construction, the matrices $ (\g_m )_{\a  \b} $ and $ (\g_m )^{\a  \b} $ are {\it real} 
and symmetric.

Upon dimensional reduction, the gravitational superfield splits into two superfields
\bea
&&H_{\a\bd}:=(\s^{\mun})_{\a\bd}H_{\mun}~~~\to~~~
H^{3\rm D}_{\a\b}=H_{\a\b}+\ri\ve_{\a\b}H~,~~~
H_{\a\b}:=(\g^m)_{\a\b}H_m =H_{\b\a}~.~~~
\eea
Here $H_{\a\b}$ is the three-dimensional gravitational superfield.
The gauge transformations (\ref{5.2}) turn into:
 \begin{subequations}
\bea
\d H_{\a \b} &=& \bar D_{(\a}L_{\b)}
-D_{(\a}\bar L_{\b)}~,  
\label{3D-gauge-H}\\
\d H &=& - \frac{\ri}{2}(\bar D_{\a}L^{\a}
-D^\a\bar L_{\a})~, \label{5.8b}\\
\d {\mathbb F}&=&\frac 14(D^\a\bar D^2 L_\a+\bar D_{\a}D^2 \bar L^{\a})~. 
\label{5.8c}
\eea
\end{subequations}
In three dimensions, the real linear scalar ${\mathbb F} $
can be expressed in terms of a real  unconstrained superfield $F = \bar F$, 
\bea
{\mathbb F} = \ri D^\a \bar D_\a F~, 
\eea
which is defined modulo arbitrary gauge transformations of the form
\bea
\d F=\l+\bar{\l}~,~~~~~~\DB_\a\l=0~.
\eea
This is the gauge transformation law of an Abelian vector  multiplet,
with $\mathbb F$ being the gauge-invariant field strength.\footnote{Given a vector multiplet,
we always use blackboard bold style to denote its gauge invariant field strength (e.g. $\mathbb F$)
and italic style to denote the gauge prepotential (e.g.  $F$). The same capital Latin letter is used for both 
the field strength and gauge prepotential.}
The supergravity gauge transformation
(\ref{5.8c}) implies that 
\bea
\d F =  \frac{\ri}{2} (\bar D_\a L^\a - D^\a \bar L_\a )~.
\eea

Dimensionally reducing the action (\ref{II-4D}) gives
\bea
S^{({\rm II})}_{\rm 3D} [H_{\a\b},H, {\mathbb F}] &=& \int \rd^3x\rd^4\q \,
\Big\{
-\frac{1}{16}H^{\a\b}D^\g\DB^2 D_\g H_{\a\b}
-\frac{1}{4}(\pa_{\a\b}H^{\a\b})^2
+\frac{1}{16}([D_\a,\DB_\b]  H^{\a\b})^2
\non\\
&&~~~~~~~~~~~
+\frac{1}{4}{\mathbb H}[D_\a,\DB_\b]  H^{\a\b}
+{\mathbb F}{\mathbb H}
+\hf {\mathbb F}{[}D_\a,\DB_\b{]} H^{\a\b}
+\frac{3}{2}{\mathbb F}^2
\Big\}~.
\label{type-II-3D-0}
\eea
Here 
${\mathbb H}$ denotes the real linear superfield
\bea
{\mathbb H}:=\ri D^\a\DB_\a H~,\qquad D^2{\mathbb H}=\DB^2{\mathbb H}=0~.
\eea
We see that $H$ appears in the action only through its gauge invariant field strength $\mathbb H$. 
Thus dimensional reduction provides us with a bonus gauge symmetry.
The supergravity gauge transformation
(\ref{5.8b}) leads to 
\bea
\d{\mathbb H}=-\frac 14(D^\a\bar D^2 L_\a+\bar D_{\a}D^2 \bar L^{\a})~.
\label{5.15}
\eea

It is useful to introduce a new parametrization for the real linear superfields:
\bea
{\mathbb G} := {\mathbb H} +2{\mathbb F}~, \qquad {\mathbb S} := {\mathbb H} +{\mathbb F}~.
\eea
As follows from (\ref{5.8c}) and (\ref{5.15}),  the superfield 
 $\mathbb S$ is invariant under the supergravity gauge transformations, 
 \bea
 \d \mathbb S =0~.
 \eea
In terms of the real linear superfields introduced, the action (\ref{type-II-3D-0}) becomes
\bea
S^{({\rm II})}_{\rm 3D} [H_{\a\b},H, {\mathbb F}] &=&
\cS^{\rm II}[H_{\a\b},{\mathbb G}]
- \hf  \int \rd^3x\rd^4\q \,{\mathbb S}^2~,
\label{5.17}
\eea
where
\bea
\cS^{\rm II}[H_{\a\b},{\mathbb G}]&=&
\int \rd^3x\rd^4\q \,
\Big\{
-\frac{1}{16}H^{\a\b}D^\g\DB^2 D_\g H_{\a\b}
-\frac{1}{4}(\pa_{\a\b}H^{\a\b})^2
+\frac{1}{16}([D_\a,\DB_\b]  H^{\a\b})^2
\non\\
&&~~
+\frac{1}{4}{\mathbb G}[D_\a,\DB_\b]  H^{\a\b}
+ \hf {\mathbb G}^2
\Big\} ~.
\label{II-3D-H}
\eea
The second term in (\ref{5.17}) describes a decoupled $\cN=2$ vector multiplet.
Therefore, the action (\ref{II-3D-H}) describes linearized $\cN=2$ supergravity.
It is invariant under the supergravity gauge transformations (\ref{3D-gauge-H}) and
\bea
\d{\mathbb G}=\frac 14(D^\a\bar D^2 L_\a+\bar D_{\a}D^2 \bar L^{\a})~.
\label{5.20}
\eea

The properties of the real linear compensator $\mathbb G$ are identical to those of $\mathbb F$.
We can introduce a real gauge prepotential $G=\bar G$ such that 
\bea
{\mathbb G} = \ri D^\a \bar D_\a G~.
\eea
The supergravity gauge transformation (\ref{5.20}) is equivalent to 
\bea
\d G =  \frac{\ri}{2} (\bar D_\a L^\a - D^\a \bar L_\a )~.
\eea

\subsection{Type I  minimal supergravity}

The supergravity action (\ref{II-3D-H}) possesses a dual formulation.
To construct it, we consider the first-order model
\bea
S^{\rm II \hookrightarrow I}&=& \int \rd^3x\rd^4\q \,
\Big\{
-\frac{1}{16}H^{\a\b}D^\g\DB^2 D_\g H_{\a\b}
-\frac{1}{4}(\pa_{\a\b}H^{\a\b})^2
+\frac{1}{16}([D_\a,\DB_\b]  H^{\a\b})^2
\non\\
&&~~
+U\Big(
\frac{1}{4}[D_\a,\DB_\b]  H^{\a\b}
-\frac{3}{2}(\s+\bar{\s})
\Big)
+ \hf U^2
\Big\} ~,
\label{duality-1}
\eea
where $U$ is unconstrained real, $U=\bar U$, and $\s$ is chiral,
\bea
\bar D_\a \s=0~.
\eea
This action proves to be  invariant under the supergravity gauge transformation (\ref{3D-gauge-H}) 
accompanied by 
\bsubeq
\bea
 \d U&=&\frac 14(D^\a\bar D^2 L_\a+\bar D_{\a}D^2 \bar L^{\a})~,\\
\d\s&=&-\frac{1}{12}\DB^2D^\a L_{\a}~.
\label{3D-gauge-sigma}
\eea
\esubeq
The superfield $\s$ act as a Lagrange multiplier for the real linear constraint.
Varying $S^{\rm II\hookrightarrow I}$ with respect to $\s$ gives
 $U={\mathbb G}$, and then the action reduces to (\ref{II-3D-H}). 
 
 On the other hand, 
if we integrate out $U$, we get the dual (Type I) action
\bea
\cS^{\rm I}[H_{\a\b},\s]&=& \int \rd^3x\rd^4\q \,
\Big\{
-\frac{1}{16}H^{\a\b}D^\g\DB^2 D_\g H_{\a\b}
-\frac{1}{4}(\pa_{\a\b}H^{\a\b})^2
+\frac{1}{32}([D_\a,\DB_\b]  H^{\a\b})^2
\non\\
&&~~
-\frac{3\ri}{4}(\s-\bar{\s})\pa_{\a\b}H^{\a\b}
-\frac{9}{4}\bar{\s}\s
\Big\}~.
\label{I-3D-H}
\eea
The supergravity gauge freedom of this action is as follows:
\begin{subequations}\label{5.28}
\bea
\d H_{\a \b} &=& \bar D_{(\a}L_{\b)}
-D_{(\a}\bar L_{\b)}~,  \\
\d\s&=&-\frac{1}{12}\DB^2D^\a L_{\a}~.
\eea
\end{subequations}

\subsection{Type III minimal supergravity}
In complete analogy to the four-dimensional case \cite{GKP},
there exist  two inequivalent ways to dualize the chiral compensator of the Type I theory 
into a real linear superfield. One of these dualities leads to the  Type II theory, while
the other produces a new dual formulation which we are going to work out below.

Let us introduce a first-order model with action
\bea
S^{\rm I \hookrightarrow III}&=& \int \rd^3x\rd^4\q \,
\Big\{
-\frac{1}{16}H^{\a\b}D^\g\DB^2 D_\g H_{\a\b}
-\frac{1}{8}(\pa_{\a\b}H^{\a\b})^2
+\frac{1}{32}([D_\a,\DB_\b]  H^{\a\b})^2
\non\\
&&~~
+ \frac{1}{4} P\Big(
\pa_{\a\b}  H^{\a\b}
+3\ri(\s-\bar{\s})
\Big)
+\frac{1}{8} P^2
\Big\}~,
\label{duality-2-2}
\eea
where $P$ is unconstrained real.
The action proves to be  invariant under 
the supergravity gauge transformations (\ref{5.28})
accompanied by 
\bea
\d P=\frac{\ri}{4}(D^\a\bar D^2 L_\a- \bar D_{\a}D^2 \bar L^{\a})~.
\eea

The model (\ref{duality-2-2}) is equivalent to Type I supergravity.
Indeed, if the field $P$ is integrated out, using its equation of motion,
 then  (\ref{duality-2-2})
 reduces to   the Type I action, eq. (\ref{I-3D-H}).
On the other hand, the equation of motion for $\s$ enforces $P$ to be linear, 
 $P= {\mathbb V}$, where $\mathbb V$ obeys the constraint 
$D^2{\mathbb V}=\DB^2{\mathbb V}=0$. As a result, the first-order action  (\ref{duality-2-2}) turns into 
 the Type III supergravity action
\bea
S^{\rm III}  [H_{\a\b},\mathbb V] &=& \int \rd^3x\rd^4\q \,
\Big\{
-\frac{1}{16}H^{\a\b}D^\g\DB^2 D_\g H_{\a\b}
-\frac{1}{8}(\pa_{\a\b}H^{\a\b})^2
+\frac{1}{32}([D_\a,\DB_\b]  H^{\a\b})^2
\non\\
&&~~
+\frac{1}{4}{\mathbb{V}}\pa_{\a\b}  H^{\a\b}
+\frac{1}{8}  {\mathbb{V}}^2
\Big\}~.
\label{III}
\eea
The corresponding gauge freedom is as follows:
\begin{subequations}
\label{5.32}
\bea
\d H_{\a \b} &=& \bar D_{(\a}L_{\b)}
-D_{(\a}\bar L_{\b)}~,  \\
\d{\mathbb V}&=&
\frac{\ri}{4} (D^\a\bar D^2 L_\a- \bar D_{\a}D^2 \bar L^{\a})~.
\label{5.32b}
\eea
\end{subequations}

Associated with the real linear scalar $\mathbb V$ is 
a real unconstrained prepotential $V=\bar V$ which is introduced by the standard  rule
\bea
{\mathbb V} = \ri D^\a \bar D_\a V~.
\eea
The supergravity gauge transformation (\ref{5.32b}) is equivalent to 
\bea
\d V = - \frac{1}{2} (\bar D_\a L^\a + D^\a \bar L_\a )~.
\eea


\subsection{Non-minimal supergravity}

The chiral compensator of the 
linearized Type I minimal supergravity, eq. (\ref{I-3D-H}),  
 can be dualized into a complex linear superfield. 
The resulting theory, which is derived below,  
describes linearized non-minimal supergravity in three dimensions.

To work out the action for linearized non-minimal supergravity, 
we introduce the following first-order action:
\bea
S^{\rm I \hookrightarrow NM}&=&
  \int \rd^3x\rd^4\q 
\Big\{
-\frac{1}{16}H^{\a\b}D^\g\DB^2 D_\g H_{\a\b}
-\frac{1}{4}(\pa_{\a\b}H^{\a\b})^2
+\frac{1}{32}([D_\a,\DB_\b]  H^{\a\b})^2
\non\\
&&
-\frac{3\ri}{2}(C-\bar{C})(\pa^{\a\b}H_{\a\b})
-\frac{3}{8}(C+\bar{C})([D^\a, \DB^\b ]H_{\a\b})
+\frac{9(1-2w)}{8}(C^2+\bar{C}^2)
\non\\
&&
-\frac{9}{4}\bar{C}C
+3C \S
+3\bar{C}\bar{\S}
\Big\}
~.
\label{I-NM}
\eea
Here $C$ is an unconstrained complex superfield, and $\S$ a complex linear superfield under the constraint
\bea
\DB^2\S=0~.
\label{CL-constraint}
\eea
The action (\ref{I-NM}) proves to be invariant under  the supergravity gauge transformation
(\ref{3D-gauge-H}) accompanied with  the following 
variations of $C$ and $\S$:
\bsubeq
\bea
\d C&=&-\frac{1}{12}\DB^2D^\a L_\a~,
\\
\d \S&=&-\frac{w+1}{8}\DB^2D^\a L_\a-\frac{1}{4}\DB_\a D^2\bar{L}^\a~.
\label{gauge-transf-NM}
\eea
\esubeq
The model (\ref{I-NM}) is equivalent to Type I supergravity.
Indeed, the equation of motion for $\S$ enforces the field $C$ to be chiral,  $\DB_\a C=0$.
After that, upon re-labelling 
$C=\s$, the action (\ref{I-NM}) reduces to (\ref{I-3D-H}).
On the other hand, when $w\ne 0,1$, one can use the equations of motion for $C$ and $\bar{C}$
in order to algebraically express these fields in terms of the other dynamical variables  in (\ref{I-NM}).
This yields  the dual non-minimal supergravity model
\bea
&&\cS^{\rm NM}[H_{\a\b},\S]=
  \int \rd^3x\rd^4\q 
\Big\{
-\frac{1}{16}H^{\a\b}D^\g\DB^2 D_\g H_{\a\b}
+\frac{w+1}{32w}([D_\a,\DB_\b]  H^{\a\b})^2
\non\\
&&~~~~~~
-\frac{w+1}{4(w-1)}(\pa_{\a\b}H^{\a\b})^2
-\frac{\ri}{(w-1)}(\S-\bar{\S})\pa^{\a\b}H_{\a\b}
-\frac{1}{4w}(\S+\bar{\S})[D^\a, \DB^\b ]H_{\a\b}
\non\\
&&~~~~~~
-\frac{1}{w(w-1)}\S\bar{\S}
+\frac{2w-1}{2w(w-1)}(\S^2+\bar{\S}^2)
\Big\}
~.~~~~~~~~
\label{3D-NM}
\eea
The corresponding gauge freedom is as follows:
\begin{subequations}
\bea
\d H_{\a \b} &=& \bar D_{(\a}L_{\b)}
-D_{(\a}\bar L_{\b)}~, \label{7.40a} \\
\d \S&=&-\frac{w+1}{8}\DB^2D^\a L_\a-\frac{1}{4}\DB_\a D^2\bar{L}^\a~.
\label{7.40b}
\eea
\end{subequations}
By construction, the action (\ref{3D-NM}) is not defined for  $w= 0,1$.
The constraint (\ref{CL-constraint}) can be solved in terms of an unconstrained prepotential 
$\bar \J^\a$, 
\bea
\S = \bar D_\a \bar \J^\a ~,
\eea
defined modulo arbitrary gauge transformations of the form 
\be
\d \bar \J^\a = \bar D_\b \bar \r^{ \a \b} ~, \qquad \bar \r^{  \b \a } = \bar \r^{ \a \b}~.
\label{CM-gauge-tran}
\ee

The parameter $w$ corresponds to the one introduced in section 
(\ref{NM-sugra-NL}).
The gauge transformation (\ref{7.40b}) and its  $w$ dependence
can in fact be inferred in few simple steps.
First, note that in four-dimensions
the non-minimal supergravity gauge transformations for the complex linear 
compensator are given by  \cite{Ideas}
\bea
\d\S=-\frac{1}{4}\frac{n+1}{3n+1}\DB^2D^\a L_\a-\frac{1}{4}\DB_\ad D^2\bar{L}^\ad~,
\label{77.43}
\eea
and are parametrized by the parameter $n$.
Then, dimensionally reduce the previous transformations to 3D and use (\ref{n--w}) to obtain 
(\ref{gauge-transf-NM}).
Once the gauge transformations are determined, the first-order action (\ref{I-NM}) 
is uniquely restored from  the requirement of its gauge invariance.

As in four dimensions (see the Appendix), 
there is a natural freedom to perform a field redefinition of $\S$ of the form 
\bea
\S ~\to ~ \S + \k \bar D_\a D_\b H^{\a\b} ~, 
\eea
with $\k$ a constant  parameter which we choose (for simplicity) to be real.  
Such a field redefinition will modify the transformation law  (\ref{7.40b}) to the form
\bea
\d \S&=&-\frac{w+1}{8}\DB^2D^\a L_\a 
-\frac{1}{4}(1-3\k) \DB_\a D^2\bar{L}^\a + \k \bar D_\a D_\b  \bar D^{(\a}L^{\b)}~.
\label{7.40b-mod}
\eea


\section{Variant supercurrents in Minkowski space}
\setcounter{equation}{0}

This section is devoted to the study of general 3D $\cN=2$ supercurrent multiplets
in Minkowski space. The general 4D $\cN=1$ supercurrents in Minkowski space
are reviewed in the Appendix.

\subsection{Supercurrents associated with off-shell supergravity}

Using the explicit structure of the three minimal actions for linearized  supergravity constructed 
above, we can derive the most general 3D $\cN=2$ supercurrent multiplet in complete analogy 
with the four-dimensional analysis given in \cite{K-var,K-Noether}.
This general procedure leads to the following  conservation equation: 
\bea 
{\bar D}^\b J_{\ab} = D_\a X + {\bar D}_\a ( \mathbb Y + \ri \mathbb Z ) ~,
\label{6.41}
\eea
where $J_{\a\b}= J_{\b\a}=\overline{J_{\a\b}} $ is the supercurrent,
and the trace multiplets $X$, $\mathbb Y$ and $\mathbb Z$ are constrained 
as follows:
\begin{subequations} \label{6.42}
\bea
{\bar D}_\a X&=& 0~, \label{6.42a}\\
{\bar D}^2 {\mathbb Y} &=& 0~, \qquad \bar{\mathbb Y} = \mathbb Y~,
\label{6.42b}\ \\
{\bar D}^2 {\mathbb Z} &=& 0~, \qquad \bar{\mathbb Z} = \mathbb Z~.
\label{6.42c}\
\eea
\end{subequations}
A 3D extension of the superfield Noether procedure \cite{MSW} 
can be argued to lead to the same $16+16$ supercurrent.
This multiplet is decomposable and can be viewed as a superposition of 
the supercurrent multiplets associated with the three minimal supergravity versions. 
The choice $\mathbb Y = \mathbb Z =0$ corresponds to the Ferrara-Zumino (FZ) multiplet 
associated with the Type I supergravity. Choosing $X= \mathbb Y =0$ gives the so-called 
$\cR$-multiplet associated with the Type II supergravity. Finally, the option $X=\mathbb Z =0$ 
corresponds to the supercurrent associated with the Type III supergravity model. Of course, 
there remains one more possibility,  $X= \mathbb Y = \mathbb Z =0$,  which holds for 
all $\cN=2$ superconformal field theories.

Given a chiral spinor superfield $\eta_\a$, such that ${\bar D}_\b \eta_\a =0$, it can always be represented 
in the form $\eta_\a = {\bar D}_\a ( \mathbb Y + \ri \mathbb Z ) $, for some real linear superfields $ \mathbb Y $ and $ \mathbb Z $ defined modulo constant shifts. 

As an example, let us derive the $\cR$-multiplet. For this we add source terms to the Type II 
action (\ref{II-3D-H})
\bea
\cS^{\rm II}[H_{\a\b},{\mathbb G}] - \hf \int \rd^3x\rd^4\q \,H^{\a\b}\cR_{\a\b} 
+ \int \rd^3x\rd^4\q \, G \mathbb Z~.
\eea
This action should preserve the vector-multiplet gauge freedom
\bea
\d G = \l + \bar \l ~, \qquad \bar D_\a \l=0
\eea
which demands the source $\mathbb Z$ to be real linear, eq. (\ref{6.42c}).
The action should also respect the linearized supergravity gauge freedom
\bea
\d H_{\a \b} &=& \bar D_{(\a}L_{\b)} -  D_{(\a} \bar L_{\b)} 
~, \qquad
\d G =  \frac{\ri}{2} (\bar D_\a L^\a - D^\a \bar L_\a )~.
\eea
This requires the sources to obey the conservation equation 
\bea 
{\bar D}^\b \cR_{\ab} =\ri  {\bar D}_\a  \mathbb Z  ~.
\eea

The $16+16$  supercurrent multiplet, eq. (\ref{6.41}),  can be modified by an improvement transformation 
of the form
\begin{subequations} \label{7.49}
\bea
 J_{\a\b} ~& \longrightarrow & ~ J_{\a\b} + D_{(\a} \bar \U_{\b)} - \bar D_{(\a}\U_{\b)}~, 
 \label{improv-111}\\
 X ~& \longrightarrow & ~ X+\hf \bar D_\a \bar \U^\a ~, 
  \label{improv-222}\\
 {\mathbb Y} ~& \longrightarrow & ~ {\mathbb Y} -\hf (D^\a \bar \U_\a +\bar D_\a \U^\a)~,
  \label{improv-333}\\
  {\mathbb Z} ~& \longrightarrow & ~ {\mathbb Z} +\ri  \,(\bar D_\a \U^\a -  D^\a \bar \U_\a)~,
   \label{improv-444}
\eea
\end{subequations}
where the spinor superfield $\U_\a$ is constrained by 
\bea
D_{(\a } \U_{\b)} =0 ~& \longrightarrow & ~ D^2 \U_\a =0~.
\eea
This constraint can locally be solved by 
\bea
\U_\a =D_\a (V +\ri \,U ) ~,\qquad \bar V =V ~, \quad \bar U =U ~,
\label{7.51}
\eea
where the scalars $V$ and $U$ are defined modulo a local shift
\bea
V +\ri \,U  ~& \longrightarrow & ~  V +\ri \,U + \bar \l ~, \qquad D_\a \bar \l =0~,
\eea
with an arbitrary chiral superfield $\l$. If $\U_\a$ is globally given by (\ref{7.51}), 
for  well defined  operators $V$ and $U$,  
then the improvement transformation  (\ref{7.49}) takes the form
\begin{subequations}\label{3D_IT}
\bea
 J_{\a\b} ~& \longrightarrow & ~ J_{\a\b} +  \big[ D_{ (\a } , \bar D_{\b )} \big] V -2 \pa_{\a\b}U~, \\
X ~& \longrightarrow & ~ X +\hf {\bar D}^2 (V-\ri U) ~, \\
{\mathbb Y} ~& \longrightarrow & ~ {\mathbb Y} + \ri D^\a{\bar D}_\a U~, \\
{\mathbb Z} ~& \longrightarrow & ~ {\mathbb Z} - 2\ri  D^\a {\bar D}_\a V~.
\eea
\end{subequations}
This is a 3D analogue of the improvement transformation given in \cite{K-Noether}.

As mentioned before, the supercurrent (\ref{6.41}) encodes information about the three minimal 
supergravity versions. We should also consider a supercurrent associated with the non-minimal 
model (\ref{3D-NM}) for linearized supergravity. Direct calculations lead to 
the non-minimal supercurrent 
\bea
{\bar D}^\b J_{\a\b} = 
- \frac{w+1}{4} D_\a {\bar D}_\b {\bar \z}^\b
-\hf {\bar D}^2 \z_\a 
 ~,
 \label{non-min_supercurrent1}
\eea
where the trace multiplet $\z_\a$ is constrained by 
\bea
D_{(\a} \z_{\b)} =0~.
\label{long_lin}
\eea
This constraint is required by the gauge invariance (\ref{CM-gauge-tran}).
The conservation law (\ref{non-min_supercurrent1})
can be rewritten in the form (\ref{6.41}) if we identify
\bea
X = - \frac{w+1}{4} {\bar D}_\a {\bar \z}^\a~, \qquad
 \mathbb Y 
 =  \hf ( D^\a \bar \z_\a + {\bar D}_\a \z^\a)~, \qquad 
 \mathbb Z = \frac{\ri}{2} ( D^\a \bar \z_\a - {\bar D}_\a \z^\a)~.~~~
\eea
As follows from (\ref{7.49}), it is always possible 
to improve at least one of  $ \mathbb Y$ and $\mathbb Z$ to zero.

It is also instructive to construct the non-minimal supercurrent 
associated with the modified transformation law (\ref{7.40b-mod}). It is 
\bea
{\bar D}^\b J_{\a\b} = 
\Big(\k - \frac{w+1}{4} \Big) D_\a {\bar D}_\b {\bar \z}^\b
+ \bar D_\a \Big\{ \k D^\b \bar \z_\b   +(1-3\k) \bar D_\b \z^\b \Big\} ~,
\eea
with $\z_\a$ constrained as in eq. (\ref{long_lin}).
This conservation law can be rewritten in the form (\ref{6.41}) provided we identify
\begin{subequations}
\bea
X&=& \Big(\k - \frac{w+1}{4} \Big)  {\bar D}_\b {\bar \z}^\b~, \\
\mathbb Y &=& \hf (1-2\k) ( D^\a \bar \z_\a + {\bar D}_\a \z^\a)~, \\
 \mathbb Z &=&  \frac{\ri}{2} (1-4\k) ( D^\a \bar \z_\a - {\bar D}_\a \z^\a)~.
\eea
\end{subequations}
It follows from these expressions that one of the three trace multiplets 
$X$, $\mathbb Y$ and $\mathbb Z$ can be set to zero by appropriately choosing  
the deformation parameter $\k$. 

Our consideration shows that non-minimal supergravity does not lead to a more general supercurrent 
than the one defined by eq. (\ref{6.41}). The reason for this  is that the non-minimal action (\ref{3D-NM}) 
can be represented as a linear combination of the three minimal actions (\ref{II-3D-H}),
(\ref{I-3D-H}) and (\ref{III}), 
\bea
\cS^{\rm NM}[H_{\a\b},\S]= a_{\rm I} \cS^{\rm I}[H_{\a\b},\s] + a_{\rm II}  \cS^{\rm II }[H_{\a\b},\mathbb G]
+ a_{\rm III}  \cS^{\rm III }[H_{\a\b},\mathbb V]~,  
\eea
for  some real parameters $a$'s such that $a_{\rm I} + a_{\rm II} + a_{\rm III} =1$, 
provided the complex linear compensator is represented as $\S = \a\, \s +\b\,  \mathbb G + \ri  \,\g \, \mathbb V$, 
with constant real coefficients $\a, \b$ and $\g$.
The derivation of this result is completely similar to the four-dimensional analysis given in \cite{GKP}.

\subsection{The S-multiplet}

The $\cS$-multiplet of Dumitrescu and Seiberg \cite{DS}, 
$\cS_{\a\b} =\cS_{\b\a} =  \overline{\cS_{\a\b}}$, 
obeys the conservation equation 
\bea
{\bar D}^\b \cS_{\a\b} = \c_\a +\cY_\a ~, 
\label{3D_S-multiplet}
\eea
where the trace multiplets $\c_\a$ and $\cY_\a$ are constrained by 
\begin{subequations}
\bea
{\bar D}_\a \c_\b &=&\hf C \ve_{\a\b} ~, \qquad D^\a \c_\a = {\bar D}_\a {\bar \c}^\a~, \\
D_{(\a} \cY_{\b )} &=&0~, \qquad {\bar D}^\a \cY_\a = -C~,
\eea
\end{subequations}
with $C$ a complex constant. Our goal is to compare the $\cS$-multiplet to the most general supercurrent 
(\ref{6.41}) derived from  off-shell supergravity.
It should be pointed out that the parameter $C$ is non-vanishing only in the presence of brane currents
\cite{DS}. Since in this paper we are interested in those rigid supersymmetric theories that can be coupled
to supergravity, we are forced  (i) to set $C=0$ and (ii) to restrict $\cY_\a$ to have the form 
\bea
\cY_\a = D_\a X~, \qquad \bar D_\a X =0~.
\eea
As a result, the $\cS$-multiplet turns into  (\ref{6.41}) with 
\bea
\mathbb Y =0~, \qquad \ri \DB_\a \mathbb Z = \c_\a~.
\eea
When $\mathbb Y =0$, the improvement transformation (\ref{7.49}) 
is generated by a superfield  $\U_\a$ constrained by \cite{DS}
\bea
D_{(\a} \U_{\b)} =0~, \qquad D^\a \bar \U_\a +\bar D_\a \U^\a =0~.
\eea

A remarkable result of Dumitrescu and Seiberg\footnote{This result was  actually 
derived in  \cite{DS} for the four-dimensional case. However, their argument can be easily 
extended to three dimensions.} \cite{DS}
is that the trace multiplet $\mathbb Y$ can always be improved 
to zero. Although their proof is based on some nontrivial assumptions,  the outcome proves to be correct 
for all known supersymmetric theories. This result has in fact a natural justification 
from the supergravity point of view, as first discussed in four dimensions \cite{K-var}. 
The point is that the $\mathbb Y$ multiplet 
is associated with the Type III minimal supergravity which is known only at the linearized 
level and does not have a nonlinear extension.  It is therefore to be expected that matter couplings to 
this supergravity formulation should be impossible.

\subsection{Examples of supercurrents}

We now give several examples of $\cN=2$ supercurrents in three dimensions.
Modulo an improvement transformation, it holds that $\mathbb Y=0$
for all models to be considered.  
Our first example is the most general supersymmetric nonlinear $\s$-model
\begin{align}
S = \int {\rm d}^3x {\rm d}^4 \q 
\,
K(\vf^I,\bar{\vf}^{\bar{J}})
+\Big{\{}
 \int {\rm d}^3x {\rm d}^2 \q  \,
W(\vf^I)
+{\rm c.c.} \Big{\}}~.
\end{align}
The corresponding supercurrent multiplet is 
\bea
\cS_{\a\b}&=&{2} K_{I\bar{J}} D_{(\a}\f^I \, \DB_{\b)}\bar \f^{\bar{J}}~, \qquad
{\mathbb Z} =- {\ri}D^{\a} {\bar D}_\a K~, \qquad
X=4W~.
\label{7.67}
\eea
This  is a 3D analogue of the 4D $\cN=1$ $\cS$-multiplet given in \cite{KS2}.
All the operators in (\ref{7.67}) are invariant under arbitrary K\"ahler transformations.

An interesting subclass of nonlinear $\s$-models is the case in which the action is invariant under
U(1)$_R$ symmetry. The $R$-symmetric K\"ahler potential $K(\phi^I,\bar{\phi}^{\bar{J}})$ 
and the superpotential $W(\phi^I)$ are respectively constrained by the equations 
(\ref{K-R-symmetric}) and (\ref{W-R-symmetric}) where the chiral superfields have 
U(1)$_R$ charges $\cJ\phi^I=-r_I\phi^I$, eq. (\ref{charges-phi-R-system}).
The equations of motion for $\f^I$ are
\bea
\DB^2 K_I&=&4W_I
~.
\eea
This equation, together with (\ref{K-R-symmetric})--(\ref{W-R-symmetric}), imply
that on-shell the superpotential $W$ admit a real prepotential $\cV$ defined by
\bea
W&=&
\frac{1}{16}\DB^2\cV
~,~~~~~~
\cV:=2\sum_Ir_I\f^I K_I=2\sum_{\bar{I}}r_I\bar{\f}^{\bar{I}}K_{\bar{I}}
~.
\eea
For the $R$-symmetric $\sigma$-model the supertrace multiplet $X$ in the supercurrent (\ref{7.67})
simplifies and takes the form $X=\frac{1}{4}\DB^2\cV$. It is clear that, by using the improvement 
transformations (\ref{improv-111})--(\ref{improv-444}), we can set to zero either ${\mathbb Z}$ or 
$X$.
In fact, by applying the improvement transformation (\ref{improv-111})--(\ref{improv-444}) 
with  $V=-\hf K$ and $U=0$ to  the supercurrent (\ref{7.67})
we obtain the FZ multiplet
 \bea
\cJ^{\rm (FZ)}_{\a\b}&=& 2 K_{I\bar{J}}D_{(\a}\phi^I \DB_{\b)}\bar{\phi}^{\bar{J}}
-\hf[D_{(\a},\DB_{\b)}]K
~,\qquad
X=\frac{1}{4}\DB^2\big(\cV-K\big)
~.
\label{R-symmetric-FZ}
\eea
On the other hand, applying the improvement transformation (\ref{3D_IT})
with  $V=\hf( K -\cV) $ and $U=0$ to the FZ multiplet leads to
  the $\cR$-multiplet 
\bea
\cR_{\a\b}&=& 2 K_{I\bar{J}} D_{(\a}\phi^I  \DB_{\b)}\bar{\phi}^{\bar{J}}
-\hf[D_{(\a},\DB_{\b)}]\cV
~,~~~
{\mathbb Z}=\ri D^\a\DB_\a \big(\cV - K\big)
~.
\label{R-symmetric-R}
\eea
The requirement that the $R$-symmetric $\sigma$-model be also
superconformal is expressed as  the condition $\cV=\cK$ \cite{KPT-MvU}.
In such a case, it follows from (\ref{R-symmetric-FZ}) and (\ref{R-symmetric-R})
that the FZ and $\cR$ multiplets  coincide.

For the next example,  consider a vector multiplet with  Chern-Simons and Fayet-Iliopoulos terms
\bea
S= \int {\rm d}^3x {\rm d}^4 \q \,
\Big\{ 
-\frac{1}{2e^2} {\mathbb G}^2 +\frac{\k}{2} G \mathbb G + \x G 
\Big\}~.
\label{VM28}
\eea
Here the parameters $\k$ and $\x$ correspond to the Chern-Simons and Fayet-Iliopoulos terms
respectively. This model is characterized by the $\cR$-multiplet \cite{DS}
\begin{subequations} \label{6.52}
\bea
J_{\a\b} &=& \frac{2}{e^2} D_{(\a} {\mathbb G } {\bar D}_{\b)} \mathbb G ~, \\
{\mathbb Z} &=& -\frac{\ri}{2e^2} D^\a\bar D_\a {\mathbb G}^2 -\x \mathbb G~, \qquad X=0~.
\eea
\end{subequations}
This is the 3D analogue of the supercurrent for the free vector multiplet model 
with a Fayet-Iliopoulos term \cite{DT,K-FI}.
The Chern-Simons coupling does not appear in (\ref{6.52}). This is due to the fact that the Chern-Simons 
term does not couple to the supergravity prepotentials. 

It is also instructive to consider an improved vector multiplet with a Chern-Simons term 
(see e.g. \cite{KLT-M})
\bea
S= \int {\rm d}^3x {\rm d}^4 \q\,
\Big\{ - \mathbb G \ln \mathbb G  +\frac{\k}{2} G \mathbb G \Big\}~.
\label{VM30}
\eea
This model is $\cN=2$ superconformal. Its supercurrent proves to be 
\bea
J_{\a\b} = \frac{2}{\mathbb G}  D_{(\a} \mathbb G \bar D_{\b)} \mathbb G 
-\hf \big[  D_{(\a} , \bar D_{\b)} \big] \mathbb G ~.
\eea
It obeys the conservation equation
\bea
\bar D^\b J_{\a\b}=0~.
\eea

The previous two models (\ref{VM28}) and (\ref{VM30})  are special cases of a 
 general system of self-interacting  Abelian vector multiplets described by the 
 gauge invariant action 
\bea
S=\int {\rm d}^3x {\rm d}^4 \q 
\,\Big\{  L({\mathbb F}^i) 
+ \hf m_{ij} F^i {\mathbb F}^j 
+ \x_{i}F^i\Big\}~.
\label{8.33VM}
\eea
Here $L$ is an arbitrary real function of 
the real linear field strengths ${\mathbb F}^i $, with $i=1,\dots, n$,
for which $F^i$ are the  gauge prepotentials, ${\mathbb F}^i=\ri D^\g\DB_\g F^i$.
The real constants $ m_{ij} =m_{ji} =(m_{ij})^*$ and $\x_i=(\x_i)^*$ are respectively
Chern-Simons and Fayet-Iliopoulos couplings.
It can be shown that the $\cR$-multiplet for this system is
\bsubeq
\bea
J_{\a\b}&=&-2L_{ij}D_{(\a} {\mathbb F}^i \DB_{\b)}{\mathbb F}^j ~, \\
{\mathbb Z}&=&
\ri D^\g\DB_{\g}\big(
{\mathbb F}^iL_{i}-L
\big)
-\x_{i}{\mathbb F}^i
~.
\eea
\esubeq
The action (\ref{8.33VM}) describes a superconformal theory if $\x_i=0$ and $L({\mathbb F})$
is a homogeneous function of first degree, ${\mathbb F}^i L_i=L$.
In this case $\mathbb Z=0$.

Finally, we consider a scalar multiplet with a real central charge 
\bea
S= \int {\rm d}^3x {\rm d}^4 \q \, \bar \F \re^{2m V_0} \F ~, \qquad 
V_0 = \ri \, \q^\a \bar \q_\a~, \qquad m= \bar m =\text{const}~.
\label{chiral-central}
\eea
Here the dynamical variable $\F$ is a chiral superfield, $\bar D_\a \F=0$.
The equations of motion imply that 
\bea
(\Box -m^2 ) \F =0~.
\eea
The supercurrent for this model is the $\cR$-multiplet \cite{DS}
\bea
J_{\a\b} = \big[  D_{(\a} , \bar D_{\b)} \big] \big( \bar \F \re^{2mV_0} \F \big)
-4\ri \, \bar \F \re^{2mV_0} 
\stackrel{\longleftrightarrow}{ \pa_{\a\b} }\F~, \qquad 
\mathbb Z = 8m \bar \F \re^{2m V_0} \F~.
\eea
Although the trace multiplet $\mathbb Z$ is linear on-shell, $\bar D^2 \mathbb Z =0$,
it cannot be represented as $\ri D^\a \bar D_\a Z $, for a well defined  operator $Z$.


\section{(1,1) AdS supersymmetry}
\setcounter{equation}{0}

In this section we study rigid supersymmetric field theories in (1,1) AdS superspace.
The geometry of this superspace is completely determined by 
the (anti-)commutation relations (\ref{AdS_(1,1)_algebra_1})--(\ref{AdS_(1,1)_algebra_2}).
As shown in subsection \ref{Type-II-cosmological}, 
the (1,1) AdS superspace originates as a solution to the equations 
of motion of the Type I minimal and the non-minimal AdS supergravity theories. 
In order to derive consistent supercurrents corresponding to (1,1) AdS supersymmetry, 
we have to compute linearized supergravity actions around the (1,1) AdS background chosen.

\subsection{Linearized minimal  supergravity}
\label{subsection9.1}
Our first task is to  derive a (1,1)  AdS extension of the linearized Type I  action in Minkowski 
superspace, eq. (\ref{I-3D-H}).
To achieve this we start with the following functional in (1,1) AdS superspace
\bea
&& \int \rd^3x\rd^4\q \,E\,
\Big\{
-\frac{1}{16}H^{\a\b}\de^\g\deb^2 \de_\g H_{\a\b}
+\frac{1}{32}([\de_\a,\deb_\b]  H^{\a\b})^2
\non\\
&&{}\qquad \qquad \qquad
-\frac{1}{4}(\de_{\a\b}H^{\a\b})^2
-\frac{3\ri}{4}(\s-\bar{\s})(\de_{\a\b}H^{\a\b})
-\frac{9}{4}\bar{\s}\s
\Big\}
\label{minimal-(1,1)-0}
~,
\eea
where the compensator $\s$ is covariantly chiral, 
\bea
\deb_\a\s=0~.
\eea
The above functional is a minimal lift of the  
Type I action (\ref{I-3D-H}) to (1,1) AdS superspace. 
The desired linearized action for minimal  AdS supergravity 
is expected to differ from 
(\ref{minimal-(1,1)-0}) by some  $\m$-dependent terms 
required to guarantee invariance under  
the linearized supergravity gauge transformations in  (1,1) AdS: 
\begin{subequations}
\bea
\d H_{\a \b} &=& \deb_{(\a}L_{\b)}
-\de_{(\a}\bar L_{\b)}~,
\label{gauge-H-1-1}
  \\
\d \s&=&-\frac{1}{12}(\deb^2-4\mu)\de^\a L_\a
~.
\label{gauge-s-1-1}
\eea
\end{subequations}

To derive the linearized  action, we compute the variation of (\ref{minimal-(1,1)-0}) under
(\ref{gauge-H-1-1}) and (\ref{gauge-s-1-1}) and then iteratively add 
certain $\m$-dependent terms in order to end up with an invariant action.
In carrying out such calculations, 
one may find useful the following identities
that derive from (\ref{AdS_(1,1)_algebra_1})--(\ref{AdS_(1,1)_algebra_2}):
\bsubeq
\bea
\de_\a\big(\de^2-4\bar{\mu}\big)&=&
4 \bar{\mu} \de^{\b}\cM_{\a\b}
~,~~~~~~
{[}\de_\a,\de^2{]}=
6\bar{\mu} \de_\a
+8\bar{\mu} \de^\b\cM_{\a\b}
~,
\\
{[}\de^\a,\deb^2{]}&=&
4\ri\de^{\a\b}\deb_\b
-6 \mu\de^\a
~,~~~
{[}\de_a,\de^2{]}=
-\ri \bar{\mu}(\g_a)_{\a\b}[\de^\a,\deb^\b]
~,
\\
{[}\de_{\a\b},\de_\g\deb^2{]}&=&
-2\ri \mu\ve_{\g(\a}\de^2\deb_{\b)}
+4 \mu\de_{\a\b}\de_\g
-4 \ri \mu\bar{\mu}\ve_{\g(\a}\deb_{\b)}
\non\\
&&
-8\ri \mu\bar{\mu}\ve_{\g(\a} \deb^{\r}\cM_{\b)\r}
+8\ri \mu\bar{\mu}\deb_{(\a}\cM_{\b)\g}
~,
\eea
\esubeq
together with their complex conjugates.
It is also important to keep in mind the rule for  integration by parts, eq. (\ref{2.12IbP}). 

The calculation procedure outlined leads to the following linearized action 
for Type I minimal AdS supergravity:
\bea
\cS^{\rm I}_{(1,1)} [H_{\a\b},\s]&=&
  \int \rd^3x\rd^4\q \,E\,
\Big\{
-\frac{1}{16}H^{\a\b}\de^\g(\deb^2-6\mu) \de_\g H_{\a\b}
\non\\
&&
-\frac{1}{4}(\de_{\a\b}H^{\a\b})^2
+\frac{1}{32}([\de_\a,\deb_\b]  H^{\a\b})^2
-\frac{3\ri}{4}(\s-\bar{\s})\de_{\a\b}H^{\a\b}
\non\\
&&
-\frac{9}{4}\bar{\s}\s
+\frac{27}{8}(\s^2+\bar{\s}^2)
\Big\}
~.~~~~~~~~~~~~
\label{11_Type-I}
\eea
Due to the identity 
\bea
\de^\a(\deb^2-6\mu)\de_\a
=
\deb_\a(\de^2-6\bar{\mu})\deb^\a
~,
\label{DDBDBD-DBDDDB}
\eea
the Lagrangian in (\ref{11_Type-I}) is manifestly real.

\subsection{Linearized non-minimal  supergravity}
\label{subsection9.2}

By analogy with 
the flat superspace case, now that we have derived the Type I action we can obtain  
a non-minimal one by dualization.
The nonlinear analysis of section \ref{NM-cosmological} tells us that this can be done only 
when $w=-1$.  This can also be immediately understood by comparing the last term in the third
line of (\ref{11_Type-I}) and the  last term in the second line of (\ref{I-NM}). Because 
the $(C^2+\bar{C}^2)$ terms in the flat first-order  action (\ref{I-NM}) and its AdS analogue 
have to be the same, it follows that only for $w=-1$ it is possible to carry out a dualization 
procedure.

A (1,1) AdS extension of the first-order action (\ref{I-NM}) with $w=-1$ should involve 
a complex unconstrained superfield $C$ and a complex linear superfield $\S$ 
constrained by 
\bea
(\deb^2-4\mu)\S=0~.
\eea
This action has the form 
\bea
&&S_{(1,1)}^{\rm I \hookrightarrow NM} [H_{\a\b},C,\S]=
  \int \rd^3x\rd^4\q \,E\,
\Big\{
-\frac{1}{16}H^{\a\b}\de^\g(\deb^2-6\mu) \de_\g H_{\a\b}
\non\\
&&~~
-\frac{1}{4}(\de_{\a\b}H^{\a\b})^2
+\frac{1}{32}([\de_\a,\deb_\b]  H^{\a\b})^2
-\frac{3\ri}{4}(C-\bar{C})\de_{\a\b}H^{\a\b}
-\frac{9}{4}\bar{C}C
+\frac{27}{8}(C^2+\bar{C}^2)
\non\\
&&~~
+3C \S
+3\bar{C}\bar{\S}
+ \frac{3}{4}C\deb^\a\de^\b H_{\a\b}
-\frac{3}{4}\bar{C}\de^\a\deb^\b H_{\a\b}
\Big\}
\label{linearizedNM}
\eea
and is invariant under the supergravity gauge transformation (\ref{gauge-H-1-1}) 
in conjunction with
\bsubeq
\bea
\d C&=&-\frac{1}{12}(\deb^2-4\mu)\de^\a L_{\a}
~,\\
\d\S&=&
-\frac{1}{4}\deb_\a(\de^2+2\bar{\mu})\bar{L}^\a
~.
\label{NM-gauge-1-1}
\eea
\esubeq
The equation of motion for $\S$ enforces the field $C$ to be chiral $\deb_\a C=0$;
with $C=\s$ the action (\ref{linearizedNM}) reduces to (\ref{11_Type-I}).
On the other hand, integrating out  $C$ and $\bar{C}$, 
we obtain the  linearized action 
for  non-minimal AdS supergravity:
\bea
\cS_{(1,1)}^{\rm NM} [H_{\a\b},\S]&=&
  \int \rd^3x\rd^4\q \,E\,
\Big\{
-\frac{1}{16}H^{\a\b}\de^\g(\deb^2-6\m) \de_\g H_{\a\b}
\non\\
&&
+\frac{\ri}{2}(\S-\bar{\S}) \de^{\a\b}H_{\a\b}
+\frac{1}{4}(\S+\bar{\S})([\de^\a,\deb^\b] H_{\a\b})
\non\\
&&
-\frac{1}{2}\bar{\S}\S
-\frac{3}{4}(\S^2+\bar{\S}^2)
\Big\}
~.~~~~~
\label{AdS-1-1-NM}
\eea
By construction, this action is invariant under the gauge transformations 
\begin{subequations}
\bea
\d H_{\a \b} &=& \deb_{(\a}L_{\b)}
-\de_{(\a}\bar L_{\b)}~, \\
\d\S&=&
-\frac{1}{4}\deb_\a(\de^2+2\bar{\mu})\bar{L}^\a ~.
\eea
\esubeq

The reader may now ask the question: 
Is it possible to dualize the linearized Type I theory in (1,1) AdS
to Type II and Type III theories?
By looking at the first-order actions (\ref{duality-1}) and (\ref{duality-2-2}), 
which are used to carry out the duality transformations in the flat case, 
it is not surprising that the answer to this question is no.
In fact, a necessary condition to perform the duality would be that the action (\ref{11_Type-I})
was a function only of the real (Type II) or imaginary (Type III) part of the chiral compensator 
$\s$. However, from the explicit form of the Type I action in (1,1) AdS, eq.  (\ref{11_Type-I}), 
it is clear that this is not the case.

\subsection{Matter couplings in (1,1) AdS superspace} 

To describe rigid supersymmetric field theories in (1,1) AdS superspace,
we need to develop a superfield description of the corresponding isometry transformations. 
The isometries are generated by (1,1) AdS Killing vector fields, $\L=\l^a\de_a+\l^\a\de_\a+\lb_\a\deb^\a$, 
which are defined to obey the master equation
\bea
\Big{[}\L+\hf\o^{ab}\cM_{ab},\de_C\Big{]}=0~.
\eea
This equation is equivalent to the relations 
\begin{subequations}
\bea
0&=&
\de_{(\a}\l_{\b)}
-\hf\o_{\a\b}
~,~~~
0=
\deb_{(\a}\l_{\b)}
+\ri\mu\l_{\a\b}
~,~~~
\de_\a\l^\a=\deb^\a\l_\a=0~,
\label{1,1-SK_1}
\\
0&=&
\de^\b\o_{\a\b}
-12\mub\l_{\a}~,
~~~
0=
\deb^\b\l_{\a\b}
+6\ri\l_{\a}~,
~~~
\de_{(\a}\l_{\b\g)}=\de_{(\a}\o_{\b\g)}=0
\label{1,1-SK_2}
\eea
\end{subequations}
and their complex conjugates. The (1,1) AdS Killing vector fields can be shown to generate the supergroup 
$\rm OSp(1|2;{\mathbb R}) \times OSp(1|2;{\mathbb R})$.

Matter couplings in  (1,1) AdS superspace are very similar to those in 4D $\cN=1$ AdS 
\cite{AJKL,FS,BKsigma}, and as such they are more restrictive than their counterparts
in Minkowski space.  As a nontrivial example, here we consider the most general 
supersymmetric nonlinear $\s$-model in (1,1) AdS superspace:
\begin{align}
S = \int {\rm d}^3x {\rm d}^4 \q 
\,E\,
\cK(\vf^I,\bar{\vf}^{\bar{J}})~.
\label{9.15}
\end{align}
The dynamical variables $\vf^I$ are covariantly chiral superfields, ${\bar \nabla}_\a \vf^I =0$,
and at the same time local complex coordinates of a complex  manifold $\cM$.
The action is invariant under (1,1) AdS isometry transformations
\bea
\d \vf^I = \L\vf^I ~.
\eea

Unlike in the Minkowski case, the action does not possess K\"ahler
invariance since 
\begin{align}\label{2.6}
\int \rd^3x\, \rd^4\theta \, E\, F (\vf)= \int \rd^3x\, \rd^2\theta \, \cE\, \mu F (\vf)  \neq 0~,
\end{align}
with $\cE$ the chiral density. Nevertheless, 
K\"ahler invariance naturally emerges if we represent the Lagrangian as
\bea
\cK(\vf,\bar{\vf})=K(\vf,\bar{\vf})+\frac{1}{\mu}W(\vf)+\frac{1}{\mub}\bar{W}(\bar{\vf})~.
\eea
Under a K\"ahler transformation, these transform as
\bea
K(\vf,\bar{\vf})\to K(\vf,\bar{\vf})+F(\vf)+\bar{F}(\bar{\vf})~,\qquad
W(\vf)\to W(\vf)-\mu F(\vf)
~.
\eea
The K\"ahler metric defined by 
\begin{align} \label{Kahler_metric}
g_{I \bar J} := \partial_I \partial_{\bar J} \cK = \partial_I \partial_{\bar J} K
\end{align}
is obviously invariant under the K\"ahler transformations.

Because of (\ref{2.6}), the Lagrangian $\cK$ in  (\ref{9.15}) should be  a globally defined function 
on the K\"ahler target space $\cM$. This immediately implies that the K\"ahler two-form, 
 $ \O=2\ri \,g_{I \bar J} \, \rd \vf^I \wedge \rd \bar \vf^{\bar J}$,  associated with 
(\ref{Kahler_metric}), is exact and hence  $\cM$ is necessarily non-compact. 
We see that the $\s$-model couplings in (1,1) AdS are more restrictive than in the Minkowski case.

\subsection{Supercurrents}

The most general supercurrent multiplet in (1,1) AdS superspace is described by 
the conservation equation
\bea
 \deb^\b J_{\a\b} = \de_\a X -\hf (\bar \de^2 +2\m ) \z_\a~,
 \label{9.21}
\eea
where $J_{\a\b} $ is the supercurrent, and $X$ and $\z_\a$ are the trace multiplets constrained by  
\begin{subequations}
\bea
\bar \de_\a X &=&0~,\\
\de_{(\a}\z_{\b)} &=&0~. \label{7.20b}
\eea
\end{subequations}
The multiplet with  $\z_\a=0$ corresponds to the Ferrara-Zumino supercurrent which is associated 
with the Type I minimal AdS supergravity. The case $X=0$ corresponds to the non-minimal AdS supergravity.
Similarly to four dimensions \cite{BKdual}, the trace multiplets in (\ref{9.21}) are defined modulo a gauge transformation of the form
\bea
X\to X+\m \L~, \qquad \z_\a \to \z_\a +  \frac{1}{4} \nabla_\a \L ~, \qquad \bar \de_\a \L =0~.
\eea
This can be used to set $X=0$.

The general supercurrent (\ref{9.21}) can be modified by an
improvement transformation 
\begin{subequations} \label{improve-nm}
\bea
J_{\a\b} &~ \longrightarrow ~ &
J_{\a\b}+\hf{[}\de_{(\a},\deb_{\b)}{]}V-2\de_{\a\b}U~, \\
X &~ \longrightarrow ~ &
X +\frac{1}{4}(\deb^2-4\mu)(V-2\ri U)~, \\
\z_\a &~ \longrightarrow ~ &\z_\a  - \de_\a(V+\ri U)~,
\eea
\end{subequations}
with $V$ and $U$ well defined  operators.

A specific feature of the $(1,1)$ AdS geometry is that the constraint (\ref{7.20b}) 
can always be solved as (compare with \cite{IS,KS94})
\bea 
\z_\a = \de_\a (V+\ri U)~, 
\eea
for well defined  operators $V$ and $U$. This property means that we can always set $\z_\a =0$
by applying a certain improvement transformation (\ref{improve-nm}).
Therefore, a Ferrara-Zumino multiplet exists for any supersymmetric field theory in the  case of $(1,1)$ AdS supersymmetry.

As an example, let us consider supercurrents for 
the supersymmetric $\s$-model (\ref{9.15}).
The non-miminal supercurrent ($X=0$) can be shown to be
\begin{subequations} 
\bea
J^{(\text{NM})}_{\a\b} &=& 2 \cK_{I \bar{J}}   \de_{(\a}\vf^I \,\deb_{\b)} \bar \vf^{\bar{J}}~, \\
\z_\a &=& - \de_\a \cK~.
\eea
\end{subequations} 
The Ferrara-Zumino multiplet is  
\begin{subequations} 
\bea
J^{(\text{FZ})}_{\a\b} &=& 2 \cK_{I \bar{J}}\de_{(\a}\vf^I \, \deb_{\b)} \bar \vf^{\bar{J}}
-\frac{1}{2} {[}\de_{(\a},\deb_{\b)}{]}\cK~, \\
X &=& -\frac{1}{4} (\deb^2-4\mu) \cK~.
\eea
\end{subequations} 
The non-minimal supercurrent looks simpler than the Ferrara-Zumino one.

\section{(2,0) AdS supersymmetry}
\setcounter{equation}{0}

In this section we study rigid supersymmetric field theories in (2,0) AdS superspace.
Its geometry is determined by 
the (anti-)commutation relations (\ref{AdS_(2,0)_algebra_1})--(\ref{AdS_(2,0)_algebra_2}).
As shown in subsection \ref{Type-II-cosmological}, this superspace 
originates as a solution to the equations 
of motion of the Type II minimal supergravity with a cosmological term.

\subsection{Linearized supergravity action}
\label{subsection10.1}

We start by deriving  a (2,0) AdS extension of the Type II action 
in Minkowski space (\ref{II-3D-H}).
To achieve this we follow the same strategy which was adopted in subsection \ref{subsection9.1}.
We start with the following functional in (2,0) AdS superspace
\bea
&& \int \rd^3x\rd^4\q \,E\,
\Big\{
-\frac{1}{16}H^{\a\b}\bfD^\g\bfDB^2 \bfD_\g H_{\a\b}
-\frac{1}{4}(\bfD_{\a\b}H^{\a\b})^2
+\frac{1}{16}([\bfD_\a,\bfDB_\b]  H^{\a\b})^2
\non\\
&& \qquad \qquad \qquad ~
+\frac{1}{4}{\mathbb G}([\bfD_\a,\bfDB_\b]  H^{\a\b})
+\frac{1}{2}{\mathbb G}^2
\Big\}~,
\label{20_type-II-0}
\eea
that reduces to the Type II action (\ref{II-3D-H}) in the flat superspace limit. 
The real linear compensator, ${\mathbb G} =\bar{\mathbb G}$, now satisfies the 
covariant constraint
\bea
\bfDB^2{\mathbb G}=0 ~.
\eea
As in Minkowski superspace, this constraint is solved in terms of a real unconstrained
prepotential $G$,
\bea
{\mathbb G}&=&\ri\bfD^\a\bfDB_\a G~,
\eea
which is defined modulo gauge shifts 
\be
\d G = \l + \bar \l~, \qquad \bar {\bf D}_\a \l=0~.
\ee 
We further postulate  linearized supergravity gauge transformations 
\begin{subequations}
\bea
\d H_{\a \b} &=& \bfDB_{(\a}L_{\b)}
-\bfD_{(\a}\bar L_{\b)}~
\label{gauge-H-2-0},
  \\
\d G&=&-\frac{\ri}{2}(\bfDB^\a L_\a-\bfD_\a \bar{L}^\a)~,
\label{gauge-G-2-0_1}
\\
\d {\mathbb G} &=&
 \frac{1}{4}(\bfD^\a\bfDB^2 L_\a+\bfDB_{\a}\bfD^2 \bar L^{\a})~.
\label{gauge-G-2-0_2}
\eea
\end{subequations}
Note that the gauge parameter $L_\a$ is charged under the U(1)$_R$:
\bea
\cJ L_\a=1~,~~~~~~
\cJ \bar{L}_\a=-1
~.
\eea

The functional (\ref{20_type-II-0}) is not
invariant under the gauge transformations (\ref{gauge-H-2-0})--(\ref{gauge-G-2-0_2}),
as can be seen using the following identities
\bsubeq
\bea
&{[}\bfD^\a,\bfDB^2{]}=
4\ri\bfD^{\a\b}\bfDB_\b
+\ri \r\bfDB^\a
-2\ri\r\bfDB^\a\cJ
-2\ri\r\bfDB_\b\cM^{\a\b}~,
\\
&\bfD_\a\bfD_\b \bfD_\g =0~,~~~
{[}\bfD_a,\bfDB^2{]}=0~,~~~
{[}\bfD_{\a\b},\bfD_\g\bfDB^2{]}=
-\hf\r\ve_{\g(\a}\bfD_{\b)}\bfDB^2
\eea
\esubeq
and their complex conjugates.
These identities can be easily derived by using the covariant derivative algebra
(\ref{AdS_(2,0)_algebra_1})--(\ref{AdS_(2,0)_algebra_2}). 
In order to get a gauge invariant action, we have to modify (\ref{20_type-II-0}) 
by adding certain $\r$-dependent terms. 
This procedure results in 
the linearized action for Type II AdS supergravity 
\bea
\cS^{\rm II}_{(2,0)}&=& \int \rd^3x\rd^4\q \,E
\Big\{
-\frac{1}{16}H^{\a\b}\bfD^\g\bfDB^2 \bfD_\g H_{\a\b}
-\frac{1}{4}(\bfD_{\a\b}H^{\a\b})^2
+\frac{1}{16}([\bfD_\a,\bfDB_\b]  H^{\a\b})^2
\non\\
&&~~
+\frac{1}{4}{\mathbb G}[\bfD_\a,\bfDB_\b]  H^{\a\b}+ \hf {\mathbb G}^2
-\frac{\ri}{4} \r H^{\a\b}\bfD^\g\bfDB_\g H_{\a\b}
+\hf \r G{\mathbb G}
\Big\}~.
\label{20_type-II}
\eea
As compared with  (\ref{20_type-II-0}), the action involves two new structures. 
The Chern-Simons 
term  coincides with that appearing in the nonlinear supergravity action (\ref{Type-II-AdS}). 
Because of its presence, the  linearized Type II AdS action cannot be dualized to 
a Type I or non-minimal model.

\subsection{Type III minimal action in (2,0) AdS superspace}
As discussed in subsection \ref{subsection9.2}, 
the Type III supergravity action (\ref{III}) cannot be lifted to (1,1) AdS superspace in a gauge invariant way.
It is quite remarkable that such an extension exists in (2,0) AdS superspace.
It has the form:
\bea
\cS_{(2,0)}^{{\rm III}}&=& \int \rd^3x\rd^4\q \,E\,
\Big\{
-\frac{1}{16}H^{\a\b}\bfD^\g\bfDB^2 \bfD_\g H_{\a\b}
-\frac{1}{8}(\bfD_{\a\b}H^{\a\b})^2
+\frac{1}{32}([\bfD_\a,\bfDB_\b]  H^{\a\b})^2
\non\\
&&
+\frac{1}{4}{\mathbb V}\bfD_{\a\b}  H^{\a\b}
+\frac{1}{8}{\mathbb V}^2
-\frac{\ri}{8}\r  H^{\a\b}\bfD^\g\bfDB_\g H_{\a\b}
-\frac{1}{4}\r V{\mathbb V}
\Big\}~.
\eea
It can be shown that this action 
is invariant under the supergravity gauge transformations\begin{subequations}
\bea
\d H_{\a \b} &=& \bar \bfD_{(\a}L_{\b)}
-\bfD_{(\a}\bar L_{\b)}~,  
\\
\d V&=&
\frac{1}{2} (\bfDB^\a L_\a+ \bfD_{\a}\bar L^{\a})
~,
\\
\d{\mathbb V}&=&
\frac{\ri}{4} (\bfD^\a\bar \bfD^2 L_\a- \bar \bfD_{\a}\bfD^2 \bar L^{\a})
~,~~~~~~
{\mathbb V}=\ri\bfD^\g\bfDB_\g V
~.
\eea
\end{subequations}

\subsection{Matter couplings in  (2,0) AdS superspace}

The isometries of (2,0) AdS superspace are generated by  Killing vector fields,
$\t=\t^a\bfD_a+\t^\a\bfD_\a+\bar{\t}_\a\bfDB^\a$, obeying the master equation
\bea
\Big{[}\t+\ri t\cJ+\hf t^{bc}\cM_{bc},\bfD_A\Big{]}=0~.
\eea
This is equivalent to the following equations on the components:
\bsubeq
\bea
&& \r\t_\a=\bfDB_\a t=\frac{\ri}{6} \r
\bfDB^\b\t_{\a\b}=\frac{\ri}{3}\bfDB^\b t_{\a\b}
~,
\\
&&\bfDB_\a\t_\b=
\bfD_{(\a}\t_{\b\g)}=\bfD_{(\a} t_{\b\g)}
=0~,\\
 &&
\bfD_\g\t^\g
=
-\bfDB^\g\bar{\t}_\g
=2\ri t ~,
\\
&&\bfD_{(\a}\t_{\b)}=-\bfDB_{(\a}\bar{\t}_{\b)}=\hf t_{\a\b}+ \frac{1}{4}\r\t_{\a\b} ~.
\eea
\esubeq
The (2,0) AdS Killing vector fields  prove to generate the supergroup 
$\rm OSp(2|2;{\mathbb R}) \times Sp(2,{\mathbb R}) $.
Rigid supersymmetric field theories in (2,0) AdS superspace should be invariant under 
the isometry transformations.

Matter couplings in  (2,0) AdS superspace significantly differ from those in the (1,1) case.
In particular, only $R$-invariant nonlinear  $\s$-models can be consistently defined in 
 (2,0) AdS superspace. As an example, consider the $\s$-model action
\begin{align}
S = \int {\rm d}^3x {\rm d}^4 \q \,E\,
K(\phi^I, \bar \phi^{\bar J})
+\Big{\{} \int {\rm d}^3x {\rm d}^2 \q \,\cE\,
W(\phi^I)
\,+\,{\rm c.c.}~\Big{\}} ~.
\label{10.11}
\end{align}
The dynamical variables $\f^I$ are covariantly chiral superfields, $\bar {\bf D}_\a \f^I=0$,
with definite  U(1)$_R$ charges $r_I$
\bea
&
\cJ \f^I= -r_I \f^I 
~,\qquad \qquad \text{(no sum)} 
~.
\eea
In order for the action to be  $R$-invariant, 
the K\"ahler potential $K(\f , \bar \f )$
and the superpotential $W (\f )$ should obey the equations:
\begin{subequations}
\bea
&& \sum_I r_I \f^I K_I =\sum_{\bar{I}} r_I \bar \f{}^{\bar{I}} K_{\bar{I}} \equiv \hf\cV(\f, \bar \f) ~, 
\label{10.17a} \\
&& \sum_I r_I \f^I W_I=2W~.
\eea
\end{subequations}
The action is invariant under the isometry transformations
\bea
\d \phi^I=
\big(\t+\ri t\cJ\big)\phi^I
~.
\eea
The  equations of motion 
\bea
\bfDB^2K_I & = &4W_I
\eea
imply that on-shell
\bea
&&\sum_I r_I \f^I \bfDB^2K_I
=8W~.
\eea
An important class of $\s$-models  in (2,0) AdS superspace
is specified by the conditions $r_i=0$ and $W(\vf)=0$. In this case no restrictions on 
the K\"ahler target space occur. Unlike the $\s$-models in (1,1) AdS superspace, 
compact target spaces are allowed.

Another interesting theory is a system of self-interacting 
Abelian vector multiplets described by 
real linear field strengths ${\mathbb F}^i $, with $i=1,\dots, n$.
A general gauge invariant action is
\bea
S=\int {\rm d}^3x {\rm d}^4 \q 
\,E\,\Big\{  L({\mathbb F}^i) 
+ \hf m_{ij} F^i {\mathbb F}^j 
+ \x_{i}F^i\Big\}~,
\label{general-abelian-2-0}
\eea
with $ m_{ij} =m_{ji} =(m_{ij})^*$ and $\x$ being Chern-Simons and Fayet-Iliopoulos coupling
constants 
respectively.
Here $F^i$ is the gauge prepotential for ${\mathbb F}^i$
and $L$ is an arbitrary real function of ${\mathbb F}^i $.
The isometry transformations of the scalar superfields $F^i$ and ${\mathbb F}^i$ are
\bea
\d F^i=\t F^i~,\qquad
\d {\mathbb F}^i=\t {\mathbb F}^i~.
\eea
Fayet-Iliopoulos terms are not allowed in (1,1) AdS superspace.

We conclude by presenting a (2,0) AdS extension of the action (\ref{chiral-central})
for a scalar multiplet with an Abelian central charge. Such an extension cannot be defined 
in the case of (1,1) AdS superspace in which a frozen vector multiplet with constant field strength
simply does not exist (since the conditions $(\bar \nabla^2 -4\m) \mathbb G =0$ and 
$\nabla_A \mathbb G =0$ are inconsistent). 
On the other hand, in (2,0) superspace such a frozen vector multiplet has been 
explicitly constructed in subsection \ref{subsection6.2}. It is described by a real gauge prepotential $V_0$
such that 
\bea
\ri \, \bfD^\a\bfDB_\a V_0=-2~.
\eea
In order to formulate our desired model,  it is useful to introduce gauge covariant derivatives 
\bea
\D_A= (\D_a, \D_\a , \bar \D^\a):= \bfD_A+\ri\G_A{\bZ}~,
\eea
where $\G_A $ is the gauge connection and $\bZ$ the central charge operator,  $[\bZ,\D_A]=0$.
 The (anti-)commutation relations for  the $\D$-derivatives  look like those in   
 (\ref{AdS_(1,1)_algebra_1})--(\ref{AdS_(1,1)_algebra_2}) except for the following
relation:
\bea
&\{\D_\a,\bar \D_\b\}
=
-2\ri\D_{\a\b}
-\ri\ve_{\a\b}(  \r\cJ + 2\bZ )
+\ri\r\cM_{\a\b}
~.
\eea
The model is described in terms of a gauge-covariant chiral superfield $\f$, 
$\bar \D_\a \f =0$, and its conjugate $\bar \f$, which are eigenvectors of the central charge, 
$\bZ \f = m \f$ and $\bZ \bar \f = - m \bar \f$, with $m$ a real mass parameter. 
In practice, it is useful to work in the chiral representation defined by 
\bea
\D_\a=\re^{-2V_0\bZ} \bfD_\a\re^{2V_0\bZ}~, \quad \bar{\D}_\a=\bfDB_\a~, 
\qquad \phi=\F~, \quad \bar{\phi}=\re^{-2V_0 \bZ} \Fb = \Fb \re^{2m V_0} ~,
\eea
with $\F$ an ordinary chiral superfield.
In the chiral representation, the action has the form
\bea
S= \int {\rm d}^3x {\rm d}^4 \q \, E\, \bar \phi \phi 
= \int {\rm d}^3x {\rm d}^4 \q \, E\, \bar \F \re^{2m V_0} \F~.
\label{massive-chiral-2,0}
\eea

\subsection{Supercurrents}

The most general supercurrent multiplet in (2,0) AdS superspace is characterized by 
the conservation equation
\bea
\bfDB^\b J_{\a\b}&=&{\bf D}_\a X -\ri \r \bar \L_\a  
+ \bfDB_\a (\mathbb Y +\ri \,  {\mathbb Z})~, 
\eea
where the trace multiplets obey the constraints 
\begin{subequations}
\bea
  \bfDB_{(\a} \bar \L_{\b)} =0~, \qquad  4X =   \bfDB_{\a} \bar \L^\a \quad \Longrightarrow 
  \quad  \bfDB_{\a} X=0 \\
\bar  {\mathbb Y} - {\mathbb Y} =\bar  {\mathbb Z} - {\mathbb Z} =0~, \qquad
\bfDB^2 {\mathbb Y}=\bfDB^2 {\mathbb Z}=0~.
\eea
\end{subequations}
The trace multiplets are defined modulo gauge transformations 
\bea \label{10.27}
\d \bar \L_\a =  -\frac{\ri}{\r}   \bfDB_{\a}
(\mathbb S +\ri \mathbb T)~,\qquad 
\d \mathbb Y = \mathbb S ~, \qquad 
\d  {\mathbb Z} = \mathbb T ~,
\eea
with $\mathbb S $ and $ \mathbb T$ real linear superfields,
\bea
\bar  {\mathbb S} - {\mathbb S} =\bar  {\mathbb T} - {\mathbb T} =0~, \qquad
\bfDB^2 {\mathbb S}=\bfDB^2 {\mathbb T}=0~.
\eea
This gauge freedom can be used to gauge away $\mathbb Y$ and $  {\mathbb Z}$.

The supercurrent can be modified by an improvement transformation of the form
\begin{subequations}\label{10.29}
\bea
J_{\a\b} &~\longrightarrow ~ & J_{\a\b} + \bfD_{(\a} \bar \U_{\b)} - 
\bar {\bf D}_{(\a} \U_{\b)}~, \\ 
\bar \L_\a &~\longrightarrow ~ & \bar \L_\a +2 \bar \U_\a \quad 
\Longrightarrow \quad 
X ~\longrightarrow ~ X +\hf \bfDB_\a \bar \U^\a~,\\
\mathbb Y &~\longrightarrow ~ & \mathbb Y -\hf ( \bfD^\a \bar \U_\a + \bfDB_\a \U^\a)~, \\
\mathbb Z &~\longrightarrow ~ & \mathbb Z +\ri (  \bfDB_\a \U^\a -  \bfD^\a \bar \U_\a )~,
\eea
\end{subequations}
where the parameter $\U_\a$ is constrained by 
\bea
\bfD_{(\a} \U_{\b)} =0~.
\eea
This freedom can be used to improve $\bar \L_\a$ to zero, thus resulting with the supercurrent 
 multiplet
 \bea
\bfDB^\b J_{\a\b}&=&
\bfDB_\a (\mathbb Y +\ri \,  {\mathbb Z})~
\eea
which is associated with the two linearized supergravity actions constructed in subsection 
\ref{subsection10.1}. We can still perform an improvement transformation 
generated by $\bar \L_\a = \bfDB_\a (\mathbb S +\ri \,  {\mathbb T})$, with 
$\mathbb S $ and $ \mathbb T$ real linear superfields. This transformation results in 
a non-zero $\bar \L_\a$ which can be set to zero by applying a certain transformation (\ref{10.27}).
We thus end up with the following improvement transformation 
\begin{subequations} \label{10.32}
\bea
J_{\a\b} &~\longrightarrow ~ & J_{\a\b} + [\bfD_{(\a},\bfDB_{\b)}] {\mathbb S}
+2 {\bf D}_{\a\b} \mathbb T ~,\\
{\mathbb Y}  &~\longrightarrow ~ & {\mathbb Y} - \ri \,  {\bf D}^\a \bar {\bf D}_\a {\mathbb T} 
+2\r {\mathbb T}~, \\
{\mathbb Z}  &~\longrightarrow ~ & {\mathbb Z} -2 \ri \,  {\bf D}^\a \bar {\bf D}_\a {\mathbb S} 
-2\r {\mathbb S}~.
\eea
\end{subequations}

In accordance with the analysis given in \cite{DS}, 
we expect that the trace multiplet $\mathbb Y$ can always 
be improved to zero. Thus any theory in (2,0) superspace should  
have a well defined 
$\cR$-multiplet described by
the conservation equation 
\bea
\bfDB^\b J_{\a\b}&=&\ri \,
\bfDB_\a {\mathbb Z}~, 
\eea
where the trace multiplet is constrained by 
\bea
\bar  {\mathbb Z} = {\mathbb Z} ~, \qquad \bfDB^2 {\mathbb Z}=0~.
\eea
The supercurrent can be modified by an improvement transformation of the form (\ref{10.32}) 
with $\mathbb Y =0$ and $\mathbb T =0$.

As an example, it can be shown that the nonlinear $\s$-model (\ref{10.11})
is characterized by the supercurrent 
\begin{subequations}
\bea
J_{\a\b} &=& 2K_{I \bar{J}}  \bfD_{(\a}\f^I \,\bfDB_{\b)}\bar \f^{\bar{J}}
-  \hf[\bfD_{(\a},\bfDB_{\b)}]\cV~,\\
{\mathbb Z} &=& -\ri \, {\bf D}^\a \bar {\bf D}_\a (K-\cV)~,
\eea
\end{subequations}
where $\cV$ is defined by  (\ref{10.17a}).
An interesting special case of $\s$-models is 
 $r_I =0$ and $W(\f) =0$. Then the K\"ahler potential is arbitrary.
The action is invariant under K\"ahler transformations
\begin{align}
K   \rightarrow K + F + \bar F~,
\end{align}
with  $F(\phi^I)$ an arbitrary holomorphic function.  
The supercurrent becomes
\begin{subequations}
\bea
J_{\a\b} &=& 2K_{I \bar{J}}  \bfD_{(\a}\f^I \,\bfDB_{\b)}\bar \f^{\bar{J}} ~,\\
{\mathbb Z} &=& -\ri \, {\bf D}^\a \bar {\bf D}_\a K ~.
\eea
\end{subequations}
The trace multiplet is clearly invariant under the K\"ahler transformations,  
and therefore it is a well defined  operator.

As another example, we consider the system of self-interacting 
Abelian vector multiplets described by the action (\ref{general-abelian-2-0}).
The supercurrent for this model is
\bsubeq
\bea
J_{\a\b}&=&-2L_{ij} \bfD_{(\a} {\mathbb F}^i \, \bfDB_{\b)}{\mathbb F}^j~,
\\
{\mathbb Z}&=&
\ri\bfD^\g\bfDB_{\g}\big(
{\mathbb F}^iL_{i}-L
\big)
-\x_{i}{\mathbb F}^i~.
\eea
\esubeq
This theory is superconformal if $\x_i=0 $ and ${\mathbb F}^iL_{i}=L$, in which case $\mathbb Z=0$.

We conclude with the supercurrent for the scalar multiplet model 
(\ref{massive-chiral-2,0}). It is an instructive exercise to show that the supercurrent 
is given by
\bsubeq
\bea
J_{\a\b}&=&
\big[  \D_{(\a} , \bar \D_{\b)} \big] (\fb\f)
-4\ri \, \fb
\stackrel{\leftrightarrow}{\D}_{\a\b} \f
=\big[  \bfD_{(\a} , \bfDB_{\b)} \big] (\fb\f)-4\ri \, \fb
\stackrel{\leftrightarrow}{\D}_{\a\b} \f
~,
\\
{\mathbb Z}&=&\big((5-4r)\r+8m\big)\fb\f
~.
\eea
\esubeq
Here $(-r)$ denotes the ${\rm U(1)}_R$ charge of $\f$.


\section{Conclusion}

In this paper we have elaborated on different aspects of three-dimensional $\cN=2$
supergravity in superspace. One of the goals was to understand how the (1,1) and (2,0) AdS 
supergravity theories \cite{AT} can be described using the different off-shell versions of $\cN=2$ 
supergravity which were briefly introduced in \cite{KLT-M}. 
The other goal was to understand the general structure of 3D $\cN=2$ supercurrents 
from the supergravity point of view. 

It was argued by Dumitrescu and Seiberg \cite{DS}
that their $\cS$-multiplet, eq. (\ref{3D_S-multiplet}), is the most general supercurrent in three dimensions. 
However, off-shell supergravity allows the existence of more general supercurrent 
described by eqs. (\ref{6.41}) and (\ref{6.42}). The same multiplet appears to emerge using a 3D analogue
of the superfield Noether procedure \cite{MSW}.
Making use of the observations given in \cite{DS}, we expect  that 
the trace multiplet $\mathbb Y$ in (\ref{6.41}) can always be improved to zero in the case of 
Poincar\'e supersymmetry. This reduces then the supercurrent (\ref{6.41}) to the $\cS$-multiplet.
However, we have shown that the $\cS$-multiplet does not have a natural extension to the 
(1,1) and (2,0) AdS superspaces. In this sense the 3D picture is very similar to the 
4D $\cN=1$ case studied in \cite{BKdual,BK_AdS_supercurrent}.

Recently there has been much interest in 3D new massive gravity
\cite{Bergshoeff:2009hq} and its supersymmetric extension
\cite{Andringa:2009yc,Bergshoeff:2010mf,Bergshoeff:2010ui}.
The results reported here should offer insight into the structure of such theories. 

Recently the problem of computing the partition function of gauge
theories on nontrivial   three- and four-dimensional constant-curvature backgrounds 
(mostly spheres)
has arisen  as a means to
compute observables such as   expectation values of Wilson loop  
and superconformal indices by using localization techniques (see 
\cite{Pestun} and \cite{Jafferis} and references therein).
The construction of supersymmetric theories on nontrivial backgrounds is itself 
an interesting  problem
and, as also pointed out in \cite{FS},
off-shell supergravity is a perfect setting to address
many related aspects.
In a sense this is a natural top-down approach: once  general supergravity-matter
couplings in superspace are understood, 
 applications to particular backgrounds arise just as an example.
On the other hand, supercurrents may serve as a powerful censor to indicate which 
supersymmetric theories can be lifted from flat to certain curved backgrounds.
We believe that the results of this paper
 can be extended to nontrivial 3D supersymmetric space-times
distinct  from AdS. 
Such applications in the case of $\cN=2$ supersymmetry, and extensions to the cases 
$\cN=3,4$   using the supergravity techniques of \cite{KLT-M}, will be studied in the future.
\\


\noindent
{\bf Acknowledgements:}\\
SMK is grateful to Daniel Butter for useful discussions and for reading the manuscript.
The work  of SMK  is supported in part by the Australian Research Council.
The work of GT-M is supported by the European 
Commission, Marie Curie Intra-European Fellowships under contract No.
PIEF-GA-2009-236454.

\appendix

\section{4D N = 1 supercurrents in Minkowski space}
\setcounter{equation}{0}
In this appendix we review the structure of 4D $\cN=1$ supercurrents in Minkowski space. 
The most general  supercurrent multiplet 
is described by the conservation equation  given in \cite{K-var,K-Noether}
\bea
&{\bar D}^{\ad}{J} _{\a \ad} =  D_\a X  + { \c}_\a  +{\rm i}\,\eta_\a ~, 
\label{Poincare_supercurrent} \\
&{\bar D}_\ad {\c}_\a  = {\bar D}_\ad \eta_\a= {\bar D}_\ad {X}=0~, 
\qquad D^\a {\c}_\a - {\bar D}_\ad {\bar {\c}}^\ad
=D^\a {\eta}_\a - {\bar D}_\ad {\bar {\eta}}^\ad = 0~.
\non
\eea
Here $J_{\a \ad} = {\bar J}_{\a \ad} $ denotes the supercurrent, and the chiral superfields 
$X$, $\c_\a$ and $\eta_\a$  constitute the so-called multiplet of anomalies. 
The above multiplet coincides with that derived by Magro, Sachs and Wolf \cite{MSW}
using their  superfield Noether procedure (see also \cite{Osborn})
provided  $\c_\a$  and $\eta_\a$ have the form:
\begin{subequations}
\bea
\c_\a &=& -\frac{1}{4} {\bar D}^2 D_\a F~, \qquad \bar F =F~,
\label{cs1} \\
\eta_\a &=& -\frac{1}{4} {\bar D}^2 D_\a H~, \qquad \bar H=H~.
\label{cs2}
\eea
\end{subequations}
However, the prepotentials $F$ and $H$ are not always well defined  operators, 
and in this sense the conservation law (\ref{Poincare_supercurrent}) is more general.\footnote{If
the prepotentials $F$ and $H$ are well defined  operators, then the supercurrent  (\ref{cs2})
can be improved 
to a Ferrara-Zumino multiplet \cite{K-Noether} (see below).}

Some of the superfields $X$,  $\c_\a$ and $\eta_\a$ 
are  absent for concrete models, and all of them can be chosen to vanish in the case of superconformal theories.
The three terms on the right of (\ref{Poincare_supercurrent}) emphasize the fact that there  
exist exactly three different linearized actions for minimal ($12+12$) supergravity, according to the 
classification given in \cite{GKP}, which are related by superfield duality transformations. 
The case  $\c_\a = \eta_\a =0$ 
describes the  Ferrara-Zumino multiplet \cite{FZ} which 
 corresponds to the old minimal formulation 
for $\cN=1$ supergravity \cite{old}. The choice  $X= \eta_\a =0$
corresponds to the new minimal supergravity \cite{new}; this supercurrent 
is called the $\cR$-multiplet \cite{GGRS,KS2}. 
Finally, the third choice $X=\c_\a  =0$
corresponds to the minimal  supergravity formulation
proposed in \cite{BGLP}; unlike the old minimal and the new minimal 
theories, this formulation is known at the linearized level only.

If only one of the superfields $\c_\a$, $\eta_\a$ and $X$  in  (\ref{Poincare_supercurrent})
is zero, the supercurrent multiplet describes $16+16$ components. Of the three such supercurrents 
studied in \cite{K-var},  the so-called $\cS$-multiplet, $\cS_{\a\ad}$,  
introduced earlier by Komargodski and Seiberg \cite{KS2}
is of fundamental significance (see below). It is described by the conservation equation
\bea
 {\bar D}^\ad \cS_{\a \ad} = D_\a X  + \c_\a  ~,
 \qquad   {\bar D}_\ad {X}= {\bar D}_\ad {\c}_\a  = 0~,  
\qquad D^\a {\c}_\a - {\bar D}_\ad {\bar {\c}}^\ad =0~.
\label{S-multiplet}
\eea

The supercurrent multiplet (\ref{Poincare_supercurrent}) can be modified by 
improvement transformations of the form \cite{K-Noether}: 
\begin{subequations} \label{A.4}
\bea
 J_{\a\bd} ~& \longrightarrow & ~ J_{\a\bd} +
 [D_\a , {\bar D}_\ad ] V- 2\pa_{\a\ad} U~,\\
X  ~& \longrightarrow & ~ X +\frac{1}{2}{\bar D}^2  (V -  {\rm i} \,U)~, \\
\c_\a ~& \longrightarrow & ~\c_\a + \frac{3}{2} {\bar D}^2 D_\a V~,\\
\eta_\a ~& \longrightarrow & ~\eta_\a + \frac{1}{2} {\bar D}^2 D_\a U~.
\eea
\end{subequations}
In terms of the spinor superfield $\U_\a = D_\a (V+\ri U )$,  
this improvement transformation can be rewritten as follows \cite{DS}:
\begin{subequations} \label{A.5}
\bea
 J_{\a\bd} ~& \longrightarrow & ~ J_{\a\bd} +D_\a \bar \U_\ad -  {\bar D}_\ad  \U_\a~,\\
X  ~& \longrightarrow & ~ X +\frac{1}{2}{\bar D}_\ad \bar \U^\ad ~, \\
\c_\a ~& \longrightarrow & ~\c_\a + \frac{3}{4} 
\Big({\bar D}^2 \U_\a -2 \bar D_\ad D_\a \bar \U^\ad - D_\a \bar D_\ad \bar \U^\ad \Big)~,\\
\eta_\a ~& \longrightarrow & ~\eta_\a -  \frac{\ri}{4} 
\Big({\bar D}^2 \U_\a +2 \bar D_\ad D_\a \bar \U^\ad + D_\a \bar D_\ad \bar \U^\ad \Big)
~.
\eea
\end{subequations}
This improvement transformation is also defined for a general spinor operator $\U_\a$
obeying only the constraint
\bea
D_{(\a} \U_{\b)} =0~.
\eea
In the case of the $\cS$-multiplet, $\eta_\a =0$ and the parameter $\U_\a$ in 
(\ref{A.5}) should be further constrained \cite{DS} by 
\bea
{\bar D}^2 \U_\a +2 \bar D_\ad D_\a \bar \U^\ad + D_\a \bar D_\ad \bar \U^\ad =0~.
\eea

It was argued in \cite{KS2} that the $\cS$-multiplet exists in all rigid supersymmetric theories
in Minkowski space.\footnote{In some exotic supersymmetric theories, the chiral scalar $X$ 
is not a well defined  operator \cite{DS}. 
In such case, the term $D_\a X$ in (\ref{S-multiplet}) should be replaced by  a spinor operator 
$\cY_\a$ constrained by $D_{(\a} \cY_{\b)} =0$ and $\bar D^2 \cY_\a =0$ \cite{DS}.}
A remarkable result of Dumitrescu and Seiberg \cite{DS} is that 
the trace multiplet $\eta_\a$ in (\ref{Poincare_supercurrent}) can always be improved 
to zero. Although their proof is based on some nontrivial assumptions, 
no counterexample is known.
This result has in fact a natural justification 
from the supergravity point of view, as first discussed in  \cite{K-var}. 
The point is that the trace multiplet $\eta_\a$
is associated with  the minimal  supergravity formulation
proposed in \cite{BGLP},
which is known only at the linearized 
level and does not have a nonlinear extension.  It is therefore to be expected that matter couplings to 
this supergravity formulation should be impossible.

It is instructive to consider the supercurrent associated with the non-minimal formulation 
for $\cN=1$ supergravity \cite{non-min,SG}. It is described by the conservation equation 
(see, e.g., \cite{K-var})
\bea
{\bar D}^{\ad}{J}_{\a \ad} = 
-\frac{1}{4} \frac{n+1}{3n+1} D_\a 
{\bar D}_\bd {\bar \z}^\bd -\frac{1}{4} {\bar D}^2  \z_\a
~, \qquad D_{(\a}\z_{\b )} =0~,
\eea
where $n$ is a real constant, $n\neq-1/3, 0$, parametrizing the different versions of 
non-minimal supergravity \cite{SG}. 
It should be pointed out that this conservation equation is based on the supergravity gauge transformation 
(\ref{77.43}) of the complex linear compensator $\S$. 
In non-minimal supergravity, there is a natural freedom to redefine $\S$ as 
\be
\S \to \S + {\k} \bar D_\ad D_\a H^{\a\ad}~,
\ee 
with $H_{\a\ad}$ the gravitational superfield, and $\k$ a constant parameter which
can be chosen (for simplicity) real. 
The redefined compensator transforms as 
\bea
\d\S=-\frac{1}{4}\frac{n+1}{3n+1}\DB^2D^\a L_\a+(\k -\frac{1}{4} )\DB_\ad D^2\bar{L}^\ad
-\k \bar D_\ad D^\a \bar D^\ad  L_\a~.
\eea
Adopting such a transformation law
leads to the conservation equation 
\bea
{\bar D}^{\ad}{J}_{\a \ad} = 
-\frac{1}{4} \frac{n+1}{3n+1} D_\a 
{\bar D}_\bd {\bar \z}^\bd + (\k-  \frac{1}{4} ){\bar D}^2  \z_\a
-{\k} \bar D_\bd D_\a \bar \z^\bd
~,
\eea
where the trace multiplet $\z_\a$ is again constrained by $D_{(\a}\z_{\b )} =0$.
This conservation equation can be written in the general form  (\ref{Poincare_supercurrent})
if we identify
\begin{subequations} \label{A.11}
\bea
X&=&\phantom{-} \frac{1}{4} \Big(2 \k -  \frac{n+1}{3n+1} \Big) {\bar D}_\ad {\bar \z}^\ad~,\\
\c_\a &=&\phantom{-} \frac{1}{4} (3\k -\hf ) 
\Big({\bar D}^2 \z_\a -2 \bar D_\ad D_\a \bar \z^\ad - D_\a \bar D_\ad \bar \z^\ad \Big)~,\\
\eta_\a &=& -\frac{\ri}{4} (\k-\hf) 
\Big({\bar D}^2 \z_\a + 2 \bar D_\ad D_\a \bar \z^\ad + D_\a \bar D_\ad \bar \z^\ad \Big)~.
\eea
\end{subequations}
There are two lessons we can learn from this example. First, the improvement transformation 
(\ref{A.5}) can be used to get rid of  either $\c_\a $ or $\eta_\a$.
Second, by an appropriate choice of the deformation parameter $\k$ we can set to zero one of 
the three trace multiplets $X$, $\c_\a$ and $\eta_\a$. The choice $\k = 1/6$ was made in 
{\it Superspace} and recently used in \cite{DS}.
Of course, if $\z_\a = D_\a \z$, for a well defined operator $\z$, 
then both $\c_\a$ and $\eta_\a$ can be improved to zero.


\begin{footnotesize}

\end{footnotesize}


\begin{thebibliography}{66}

\bibitem{Siegel80}
  W.~Siegel,
  ``Off-shell central charges,''
  Nucl.\ Phys.\  B {\bf 173}, 51 (1980).

\bibitem{A-GF}
L.~Alvarez-Gaum\'e and D.~Z.~Freedman,
``Potentials for the supersymmetric nonlinear sigma model,''
Commun.\ Math.\ Phys.\  {\bf 91}, 87 (1983).

\bibitem{SS}
  J.~Scherk and J.~H.~Schwarz,
  ``How to get masses from extra dimensions,''
  Nucl.\ Phys.\  B {\bf 153}, 61 (1979).

\bibitem{DJT1}
  S.~Deser, R.~Jackiw and S.~Templeton,
  ``Three-dimensional massive gauge theories,''
  Phys.\ Rev.\ Lett.\  {\bf 48}, 975 (1982).
\bibitem{DJT2}
  S.~Deser, R.~Jackiw and S.~Templeton,
  ``Topologically massive gauge theories,''
  Annals Phys.\  {\bf 140}, 372 (1982)
  [Erratum-ibid.\  {\bf 185}, 406 (1988)].
 

\bibitem{Witten}
E.~Witten,
  ``(2+1)-dimensional gravity as an exactly soluble system,''
Nucl.\ Phys.\  B {\bf 311}, 46 (1988).

\bibitem{HW}
J.~H.~Horne and E.~Witten,
``Conformal gravity in three dimensions as a gauge theory,''
Phys.\ Rev.\ Lett.\  {\bf 62}, 501 (1989).

\bibitem{vN}
  P.~van Nieuwenhuizen,
  ``D = 3 conformal supergravity and Chern-Simons terms,''
  Phys.\ Rev.\  D {\bf 32}, 872 (1985).

\bibitem{RvN}
  M.~Ro\v{c}ek and P.~van Nieuwenhuizen,
  ``${\rm N}   \geq 2$ supersymmetric Chern-Simons terms as D = 3 extended conformal
  supergravity,''
  Class.\ Quant.\ Grav.\  {\bf 3}, 43 (1986).
  
\bibitem{AT}
  A.~Ach\'ucarro and P.~K.~Townsend,
  ``A Chern-Simons action for three-dimensional anti-de Sitter supergravity
 theories,''
  Phys.\ Lett.\  B {\bf 180}, 89 (1986).

\bibitem{LR89}
  U.~Lindstr\"om and M.~Ro\v{c}ek,
  ``Superconformal gravity in three dimensions as a gauge theory,''
  Phys.\ Rev.\ Lett.\  {\bf 62}, 2905 (1989).

\bibitem{AT2}
  A.~Ach\'ucarro and P.~K.~Townsend,
  ``Extended supergravities  in d = (2+1) as Chern-Simons theories,''
  Phys.\ Lett.\  B {\bf 229}, 383 (1989).

 
\bibitem{HIPT}
  P.~S.~Howe, J.~M.~Izquierdo, G.~Papadopoulos and P.~K.~Townsend,
  ``New supergravities with central charges and Killing spinors in 2+1 dimensions,''
  Nucl.\ Phys.\  B {\bf 467}, 183 (1996)
  [arXiv:hep-th/9505032].


\bibitem{IT}
  J.~M.~Izquierdo and P.~K.~Townsend,
  ``Supersymmetric space-times in (2+1) adS supergravity models,''
  Class.\ Quant.\ Grav.\  {\bf 12}, 895 (1995)
  [arXiv:gr-qc/9501018].



\bibitem{GGRS}
S.~J.~Gates Jr., M.~T.~Grisaru, M.~Ro\v{c}ek and W.~Siegel,
{\it Superspace, or One Thousand and One Lessons in Supersymmetry},
Benjamin/Cummings (Reading, MA),  1983, hep-th/0108200.

\bibitem{NG}
  H.~Nishino and S.~J.~Gates Jr.,
  ``Chern-Simons theories with supersymmetries in three dimensions,''
  Int.\ J.\ Mod.\ Phys.\  A {\bf 8}, 3371 (1993).

\bibitem{KT}
M.~Kaku and P.~K.~Townsend,
``Poincar\'e supergravity as broken superconformal gravity,''
  Phys.\ Lett.\  B {\bf 76}, 54 (1978).
 
\bibitem{KLT-M}
 S.~M.~Kuzenko, U.~Lindstr\"om and G.~Tartaglino-Mazzucchelli,
``Off-shell supergravity-matter couplings in three dimensions,''
JHEP {\bf 1103}, 120 (2011)  [arXiv:1101.4013 [hep-th]].

\bibitem{Howe:2004ib}
P.~S.~Howe and E.~Sezgin,
``The supermembrane revisited,''
Class.\ Quant.\ Grav.\ {\bf 22} (2005) 2167
[arXiv:hep-th/0412245].

\bibitem{CGN}
  M.~Cederwall, U.~Gran and B.~E.~W.~Nilsson,
  ``D=3, N=8 conformal supergravity and the Dragon window,''
  arXiv:1103.4530 [hep-th].

\bibitem{GH}
  J.~Greitz and P.~S.~Howe,
  ``Maximal supergravity in three dimensions: supergeometry and differential forms,''
  JHEP {\bf 1107}, 071 (2011)
  [arXiv:1103.2730 [hep-th]].

\bibitem{Ideas} 
I.~L. Buchbinder and S.~M. Kuzenko, {\it Ideas and Methods of Supersymmetry and
Supergravity, Or a Walk Through Superspace}, IOP, Bristol, 1998.

\bibitem{Howe}
P.~S.~Howe,
``A superspace approach to extended conformal supergravity,''
Phys.\ Lett.\  B {\bf 100}, 389 (1981);
``Supergravity in superspace,''  Nucl.\ Phys.\  B {\bf 199}, 309 (1982).

\bibitem{old}
J.~Wess and B.~Zumino,
``Superfield Lagrangian for supergravity,''
Phys.\ Lett.\  B {\bf 74}, 51 (1978);
K.~S.~Stelle and P.~C.~West,
``Minimal auxiliary fields for supergravity,''
Phys.\ Lett.\  B {\bf 74},  330 (1978);
S.~Ferrara and P.~van Nieuwenhuizen,
``The auxiliary fields of supergravity,''
Phys.\ Lett.\  B {\bf 74}, 333 (1978).

\bibitem{WB} J.~Wess and J.~Bagger,
{\it Supersymmetry and Supergravity},
Princeton University Press, Princeton, 1992.

\bibitem{new}
V.~P.~Akulov, D.~V.~Volkov and V.~A.~Soroka,
``Generally covariant theories of gauge fields on superspace,'' 
Theor.\ Math.\ Phys.\  {\bf 31}, 285 (1977);
M.~F.~Sohnius and P.~C.~West,
``An alternative minimal off-shell version of N=1 supergravity,''
Phys.\ Lett.\  B {\bf 105}, 353 (1981).

\bibitem{non-min}
P.~Breitenlohner,
 ``A geometric interpretation of local supersymmetry,''
  Phys.\ Lett.\  B {\bf 67}, 49 (1977);
``Some invariant Lagrangians for local supersymmetry,''
Nucl.\ Phys.\ {\bf B124}, 500 (1977).

\bibitem{SG}
W.~Siegel and S.~J.~Gates Jr.
 ``Superfield supergravity,''  Nucl.\ Phys.\  B {\bf 147}, 77 (1979).

\bibitem{BKdual}
D.~Butter and S.~M.~Kuzenko,
``A dual formulation of supergravity-matter theories,''
arXiv:1106.3038 [hep-th].

\bibitem{BK_AdS_supercurrent}
D.~Butter and S.~M.~Kuzenko,
``N=2 AdS supergravity and supercurrents,''
  JHEP {\bf 1107}, 081 (2011)
  [arXiv:1104.2153 [hep-th]].

\bibitem{BKsigma}
D.~Butter and S.~M.~Kuzenko,
 ``N=2 supersymmetric sigma-models in AdS,''
arXiv:1105.3111 [hep-th];
  ``The structure of N=2 supersymmetric nonlinear sigma models in AdS4,''
  arXiv:1108.5290 [hep-th].



\bibitem{AJKL}
  A.~Adams, H.~Jockers, V.~Kumar and J.~M.~Lapan,
 ``N=1 sigma models in $AdS_4$,''
  arXiv:1104.3155 [hep-th].
 
\bibitem{FS}
G.~Festuccia and N.~Seiberg,
``Rigid supersymmetric theories in curved superspace,''
JHEP {\bf 1106}, 114 (2011)  [arXiv:1105.0689 [hep-th]].

\bibitem{FZ}
  S.~Ferrara and B.~Zumino,
``Transformation properties of the supercurrent,''
  Nucl.\ Phys.\  B {\bf 87}, 207 (1975).
  
\bibitem{OS}
V.~Ogievetsky and E.~Sokatchev,
``On vector superfield generated by supercurrent,''
Nucl.\ Phys.\  B {\bf 124}, 309 (1977).

\bibitem{FZ2}
  S.~Ferrara and B.~Zumino,
  ``Structure of conformal supergravity,''
  Nucl.\ Phys.\  B {\bf 134}, 301 (1978).

\bibitem{Siegel}
  W.~Siegel,
  ``A derivation of the supercurrent superfield,''
Harvard  preprint  HUTP-77/A089 (December, 1977). 

\bibitem{CPS}
  T.~E.~Clark, O.~Piguet and K.~Sibold,
  ``Supercurrents, renormalization and anomalies,''
  Nucl.\ Phys.\  B {\bf 143}, 445 (1978).

\bibitem{GGS}
  S.~J.~Gates Jr., M.~T.~Grisaru and W.~Siegel,
  ``Auxiliary  field anomalies,''
  Nucl.\ Phys.\  B {\bf 203}, 189 (1982).

\bibitem{Osborn}
  H.~Osborn,
  ``N = 1 superconformal symmetry in four-dimensional quantum field theory,''
  Annals Phys.\  {\bf 272}, 243 (1999)
  [arXiv:hep-th/9808041].

\bibitem{MSW}
  M.~Magro, I.~Sachs and S.~Wolf,
  ``Superfield Noether procedure,''
  Annals Phys.\  {\bf 298}, 123 (2002)
  [arXiv:hep-th/0110131].
  
\bibitem{KS-FI}
  Z.~Komargodski and N.~Seiberg,
``Comments on the Fayet-Iliopoulos term in field theory and supergravity,''
 JHEP {\bf 0906}, 007 (2009)  [arXiv:0904.1159 [hep-th]].

\bibitem{DT}
  K.~R.~Dienes and B.~Thomas,
  ``On the inconsistency of Fayet-Iliopoulos terms in supergravity theories,''
  Phys.\ Rev.\ D {\bf 81}, 065023 (2010). 

\bibitem{K-FI}
S.~M.~Kuzenko,
``The Fayet-Iliopoulos term and nonlinear self-duality,''
 Phys.\ Rev.\  D {\bf 81}, 085036 (2010)
 [arXiv:0911.5190 [hep-th]].
  
\bibitem{KS2}
Z.~Komargodski and N.~Seiberg,
``Comments on supercurrent multiplets, supersymmetric field theories and supergravity,''
JHEP {\bf 1007}, 017 (2010) [arXiv:1002.2228 [hep-th]].

\bibitem{K-var}
S.~M.~Kuzenko,
``Variant supercurrent multiplets,''
JHEP {\bf 1004}, 022 (2010)  [arXiv:1002.4932 [hep-th]].

\bibitem{Butter-FI}
  D.~Butter,
  ``Conserved supercurrents and Fayet-Iliopoulos terms in supergravity,''
  arXiv:1003.0249 [hep-th].

\bibitem{ZH}
S.~Zheng and J.~h.~Huang,
 ``Variant Supercurrents and Linearized Supergravity,''
  Class.\ Quant.\ Grav.\  {\bf 28}, 075012 (2011)
  [arXiv:1007.3092 [hep-th]].


 \bibitem{K-Noether}
 S.~M.~Kuzenko,
  ``Variant supercurrents and Noether procedure,''
  Eur.\ Phys.\ J.\  C {\bf 71}, 1513 (2011)
  [arXiv:1008.1877 [hep-th]].
  
\bibitem{DS}
  T.~T.~Dumitrescu and N.~Seiberg,
  ``Supercurrents and brane currents in diverse dimensions,''
JHEP {\bf 1107}, 095 (2011)
  [arXiv:1106.0031 [hep-th]].

\bibitem{BK_supercurrent}
  D.~Butter and S.~M.~Kuzenko,
  ``N=2 supergravity and supercurrents,''
  JHEP {\bf 1012}, 080 (2010)
  [arXiv:1011.0339 [hep-th]].

\bibitem{BTZ1}
  M.~Banados, C.~Teitelboim and J.~Zanelli,
  ``The black hole in three-dimensional space-time,''
  Phys.\ Rev.\ Lett.\  {\bf 69}, 1849 (1992)
  [arXiv:hep-th/9204099].

\bibitem{BTZ2}
  M.~Banados, M.~Henneaux, C.~Teitelboim and J.~Zanelli,
  ``Geometry of the (2+1) black hole,''
  Phys.\ Rev.\  D {\bf 48}, 1506 (1993)
  [arXiv:gr-qc/9302012].
  
\bibitem{CH}
  O.~Coussaert and M.~Henneaux,
  ``Supersymmetry of the (2+1) black holes,''
  Phys.\ Rev.\ Lett.\  {\bf 72}, 183 (1994)
  [arXiv:hep-th/9310194].

\bibitem{KMc}
  S.~M.~Kuzenko and S.~A.~McCarthy,
  ``On the component structure of N=1 supersymmetric nonlinear
  electrodynamics,''
  JHEP {\bf 0505}, 012 (2005)
  [arXiv:hep-th/0501172].

\bibitem{BM}
  I.~A.~Bandos and C.~Meliveo,
  ``Three form potential in (special) minimal supergravity superspace and
  supermembrane supercurrent,''
  arXiv:1107.3232 [hep-th].


\bibitem{DG85}
B.~B.~Deo and S.~J.~Gates Jr.,
``Comments on nonminimal N=1 scalar multiplets,''
  Nucl.\ Phys.\ B {\bf 254}, 187 (1985).


\bibitem{KTyler}
S.~M.~Kuzenko and S.~J.~Tyler,
``Complex linear superfield as a model for Goldstino,''
  JHEP {\bf 1104}, 057 (2011)
  [arXiv:1102.3042 [hep-th]].



\bibitem{GKP}
  S.~J.~Gates Jr., S.~M.~Kuzenko and J.~Phillips,
  ``The off-shell (3/2,2) supermultiplets revisited,''
  Phys.\ Lett.\  B {\bf 576}, 97 (2003)
  [arXiv:hep-th/0306288].


   
\bibitem{BGLP}
I.~L.~Buchbinder, S.~J.~Gates Jr., W.~D.~Linch and J.~Phillips,
``New 4D, N = 1 superfield theory: Model of free massive superspin-3/2  multiplet,''
Phys.\ Lett.\  B {\bf 535}, 280 (2002)
[arXiv:hep-th/0201096].
 

\bibitem{IS}
 E.~A.~Ivanov and A.~S.~Sorin,
``Superfield formulation of OSp(1,4) supersymmetry,''
J.\ Phys.\ A  {\bf 13}, 1159 (1980).


\bibitem{KS94}
  S.~M.~Kuzenko and A.~G.~Sibiryakov,
  ``Free massless higher superspin superfields on the anti-de Sitter superspace,''
Phys.\ Atom.\ Nucl.\  {\bf 57}, 1257 (1994)
  [Yad.\ Fiz.\  {\bf 57}, 1326 (1994)].



\bibitem{BILS}
I.~A.~Bandos, E.~Ivanov, J.~Lukierski and D.~Sorokin,
``On the superconformal flatness of AdS superspaces,''
JHEP {\bf 0206}, 040 (2002) [hep-th/0205104].

\bibitem{4D-N=2-conf-flat}
 S.~M.~Kuzenko and G.~Tartaglino-Mazzucchelli,
  ``Field theory in 4D N=2 conformally flat superspace,''
  JHEP {\bf 0810}, 001 (2008)
  [arXiv:0807.3368 [hep-th]].
  
\bibitem{5D-N=1-conf-flat}
S.~M.~Kuzenko and G.~Tartaglino-Mazzucchelli,
``Conformally flat supergeometry in five dimensions,''
JHEP {\bf 0806}, 097 (2008)  [arXiv:0804.1219 [hep-th]].



\bibitem{KPT-MvU}
S.~M.~Kuzenko, J.~H.~Park, G.~Tartaglino-Mazzucchelli and R.~Unge,
``Off-shell superconformal nonlinear sigma-models in three dimensions,''
JHEP {\bf 1101}, 146 (2011)
  [arXiv:1011.5727 [hep-th]].






\bibitem{Bergshoeff:2009hq}
  E.~A.~Bergshoeff, O.~Hohm and P.~K.~Townsend,
 ``Massive gravity in three dimensions,''
  Phys.\ Rev.\ Lett.\  {\bf 102}, 201301 (2009)
  [arXiv:0901.1766 [hep-th]].


\bibitem{Andringa:2009yc}
  R.~Andringa, E.~A.~Bergshoeff, M.~de Roo, O.~Hohm, E.~Sezgin and P.~K.~Townsend,
  ``Massive 3D supergravity,''
  Class.\ Quant.\ Grav.\  {\bf 27}, 025010 (2010)
  [arXiv:0907.4658 [hep-th]].

\bibitem{Bergshoeff:2010mf}
  E.~A.~Bergshoeff, O.~Hohm, J.~Rosseel, E.~Sezgin and P.~K.~Townsend,
  ``More on massive 3D supergravity,''
  arXiv:1005.3952 [hep-th].

\bibitem{Bergshoeff:2010ui}
  E.~A.~Bergshoeff, O.~Hohm, J.~Rosseel and P.~K.~Townsend,
  ``On maximal massive 3D supergravity,''
  Class.\ Quant.\ Grav.\  {\bf 27}, 235012 (2010)
  [arXiv:1007.4075 [hep-th]].


\bibitem{Pestun}
  V.~Pestun,
  ``Localization of gauge theory on a four-sphere and supersymmetric Wilson loops,''
[arXiv:0712.2824 [hep-th]]; 
``Localization of the four-dimensional N=4 SYM to a two-sphere and 1/8 BPS Wilson loops,''
[arXiv:0906.0638 [hep-th]].

\bibitem{Jafferis}
  D.~L.~Jafferis,
  ``The exact superconformal R-symmetry extremizes Z,''
arXiv:1012.3210 [hep-th].


\end{thebibliography}
\end{document}